\documentclass[prc,showkeys,showpacs,preprintnumbers,amsmath,amssymb]{revtex4}

\usepackage{amssymb}
\usepackage{graphicx,color}% Include figure files

\begin{document}

%\preprint{\hfill LBNL-}

\title{Multiple Parton Scattering in Nuclei: \\
Modified DGLAP Evolution for Fragmentation Functions}

\author{Wei-Tian Deng$^{1,2}$ and Xin-Nian Wang$^{2}$ }
 \affiliation{$^{1}$School of Physics, Shandong University, Jinan, Shandong  250100, China.\\
 $^{2}$Nuclear Science Division, MS 70R0319, Lawrence Berkeley National Laboratory, Berkeley, California 94720 }
%\author{Xin-Nian Wang}
% \affiliation{Nuclear Science Division, MS 70R0319, \\
%Lawrence Berkeley National Laboratory, Berkeley, California 94720 }

%\address[shanda]{Department of Physics, Shandong University, Jinan,
%Shandong 250100, China}
%\address[lbnl]{Nuclear Science Division, MS 70R0319,
%Lawrence Berkeley National Laboratory, Berkeley, California 94720}

\date{\today}% It is always \today, today,

\begin{abstract}
Within the framework of generalized factorization of higher-twist
contributions to semi-inclusive cross section of deeply inelastic
scattering off a large nucleus, multiple parton scattering leads to
an effective medium-modified fragmentation function and the
corresponding medium-modified DGLAP evolution equations. We extend
the study to include gluon multiple scattering and induced
quark-antiquark production via gluon fusion . We numerically solve
these medium-modified DGLAP (mDGLAP) evolution equations and study
the scale ($Q^2$), energy ($E$), length ($L$) and jet transport
parameter ($\hat q$) dependence of the modified fragmentation
functions for a jet propagating in a uniform medium with finite
length (a ``brick'' problem). We also discuss the concept of parton
energy loss within such mDGLAP evolution equations and its
connection to the modified fragmentation functions. With a realistic
Wood-Saxon nuclear geometry, we calculate the modified fragmentation
functions and compare to experimental data of DIS off large nuclei.
The extracted jet transport parameter at the center of a large
nucleus is found to be $\hat q_{0}=0.024\pm0.008$ GeV$^{2}$/fm.
\end{abstract}

\pacs{Valid PACS appear here}% PACS, the Physics and Astronomy

%\keywords{Modified splitting functions; Modified Fragmentation
%Functions; Higher-twist expansion; Energy loss}
%Use showkeys class option if keyword
\maketitle

%\begin{multicols}{2}

\section{Introduction}
Jet quenching or the suppression of large transverse momentum spectra \cite{wg90} can be used as
an effective probe of the properties of dense medium created in high-energy heavy-ion collisions.
Because of multiple scattering and induced gluon bremsstrahlung, an energetic parton propagating in
dense medium will lose a significant amount of energy \cite{Gyulassy:1993hr,Baier:1996sk,Wiedemann:2000za,
Gyulassy:2000er,Guo:2000nz,Wang:2001ifa} and therefore soften its final
fragmentation functions. These modified fragmentation functions will lead to the suppression of
large transverse momentum single hadron spectra\cite{Wang:1998bha,Wang:1998ww}, photon-hadron
correlations \cite{Wang:1996yh,Wang:1996pe,arleo,Renk:2006qg,zoww09,Qin:2009bk}
both away-side \cite{Wang:2003mm,zoww07} and same-side dihadron correlations \cite{Majumder:2004pt}
in high-energy heavy-ion collisions. Such proposed phenomena have indeed been observed in experiments  \cite{phenix,star0,star,star-gam-hadr,frantz} at the Relativistic Heavy-ion Collider (RHIC).
Phenomenological studies of the observed jet quenching
phenomena at RHIC indicate a scenario of strong interaction between energetic partons and the hot
medium with an extremely high initial parton density \cite{Wang:2003mm,Vitev:2002pf,Eskola:2004cr,Turbide:2005fk,majumder}. The same phenomena are also
predicted in deeply inelastic scattering (DIS) off large nuclei when the struck quark propagates through the
target nuclei \cite{ww02} though the extracted parton density in cold nuclei is much smaller than that
in the hot matter produced in the central $Au+Au$ collisions at RHIC.

Large transverse momentum hadrons in high-energy nucleon-nucleon collisions are produced
through hard parton scattering with large transverse momentum transfer and the subsequent
fragmentation of energetic partons into final hadrons. Because of the large transverse momentum transfer
involved the single inclusive, dihadron and gamma-hadron cross section at large transverse momentum
can be calculated within the collinear factorized parton model of perturbative QCD
(pQCD) \cite{Aurenche:1999nz,Owens:2001rr,Baer:1990ra}. Since the initially produced partons
are extremely virtual they will have to  go through a cascade of vacuum gluon bremsstrahlung before
the final hadronization at a scale $Q_{0}^{2}\sim 1$ GeV$^{2}$ below which pQCD is no longer
valid. In the leading logarithmic approximation, the transverse momenta of the emitted gluons are
ordered and therefore each successive gluon emission will contribute to one power of
a logarithmic factor $\ln (Q^2/Q_0^2)$ which will counter the small $\alpha_{s}(Q^{2})$ for large
momentum scale $Q^{2}$. The resummation of these vacuum gluon bremsstrahlung will lead
to a scale dependence of the jet fragmentation functions which can be described by
the Dokshitzer-Gribov-Lipatov-Altarelli-Parisi (DGLAP) \cite{dglap} evolution equations in pQCD.

In the presence of nuclear and hot QCD medium, the initially produced energetic partons will have to go
through multiple scattering and induced gluon bremsstrahlung as they propagate through the medium
and before hadronization. The induced gluon bremsstrahlung effectively reduces the leading parton's
energy and softens the final hadron spectra or parton fragmentation functions. Within the framework
of generalized factorization of higher-twist contribution from multiple parton scattering, one can cast the
modification of the semi-inclusive hadron cross section due to multiple parton scattering and induced
gluon bremsstrahlung in terms of the effective medium modified parton fragmentation
functions \cite{Guo:2000nz,Wang:2001ifa}. In the same spirit of the vacuum gluon
bremsstrahlung, a medium modified DGLAP (mDGLAP) equation was also derived in
 Refs.~\cite{Guo:2000nz,Wang:2001ifa} which in effect resums successive induced gluon
 bremsstrahlung due to multiple scattering. The form of the mDGLAP evolution equation is the
 same as in the vacuum except that the effective splitting functions will
 contain both vacuum bremsstrahlung and the medium induced parts which
 should contain information about the properties of the medium as probed by the energetic parton.

 Resummation of medium induced gluon bremsstrahlung in terms of a Monte Carlo simulation of
 parton shower has been carried out \cite{Armesto:2007dt, Zapp:2008af} within the framework of opacity expansion for
 induced gluon bremsstrahlung \cite{Wiedemann:2000za}, which is equivalent to solving the mDGLAP
 evolution equations. The mDGLAP evolution equations as derived in the higher-twist
 approach \cite{Guo:2000nz,Wang:2001ifa} have also been recently studied numerically \cite{Majumder:2009zu}
 for modified fragmentation function in DIS off nuclear targets as this work is being completed.

In this paper, we will extend the mDGLAP evolution equations as derived
in Refs.~\cite{Guo:2000nz,Wang:2001ifa} to include both induced gluon bremsstrahlung
and induced quark-antiquark pair production from gluon-medium interaction within the
same higher-twist expansion approach. We will then solve the coupled mDGLAP evolution
equations for gluon and quark fragmentation functions using a numerical method based on
Runge-Kutta iteration. We will first consider parton propagation in a uniform medium with finite length
and a constant gluon density (so-called ''brick'' problem). We will study in detail how the modified
fragmentation function from mDGLAP
evolution equations depend on the initial parton energy, the momentum scale, medium length
and the jet transport parameter (or gluon density). Since the concept of parton energy loss
becomes ambiguous in the process of successive gluon bremsstrahlung and quark-antiquark
pair production, we will consider the singlet quark distribution within a propagating jet with the initial
condition of a $\delta$-function $\delta (1-z)$ and a given flavor. We will study the momentum
fraction carried by the singlet quark distribution function of the fixed flavor and compare to the
quark energy loss as calculated in the traditional picture of induced gluon bremsstrahlung.
With a realistic Wood-Saxon geometry  of cold nuclear matter, we then calculate the nuclear
modified quark fragmentation functions from the mDGLAP evolution equations in the DIS off nuclei
and compare to experimental data to extract the jet quenching parameter in cold nuclear matter.

\section{Modified fragmentation functions}

\begin{figure}
  \centering
  \includegraphics[width=0.6\textwidth]{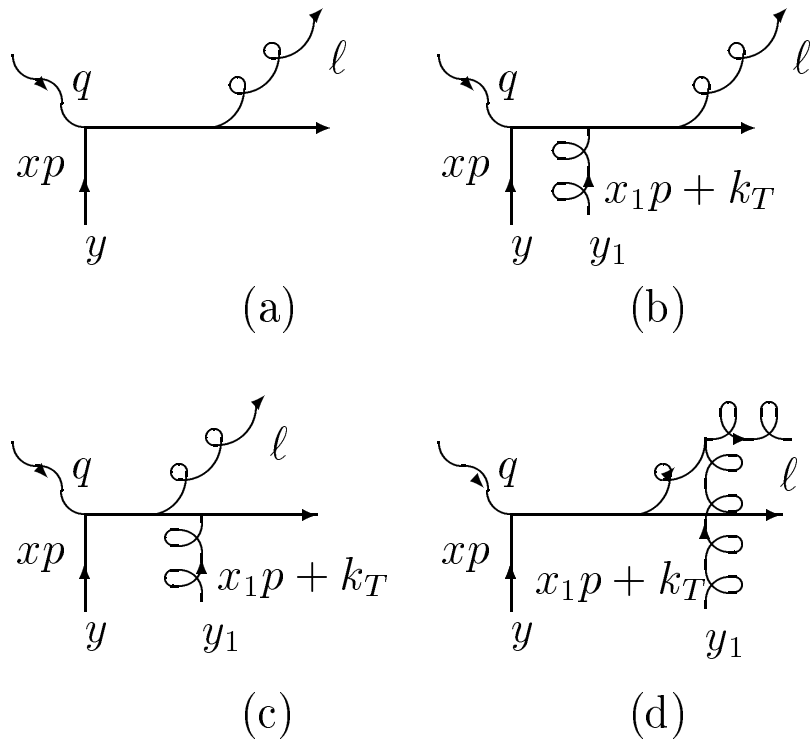}%
  \caption{Gluon radiation from a single scattering (a) and double scattering (b-d) in DIS off a nucleus.}
  \label{fig:twist_four}
\end{figure}

In DIS off a nucleus target, the struck quark via photon-nucleon interaction is likely to scatter with
another nucleon along its propagation path inside the nucleus. Such secondary quark nucleon
scattering can also be accompanied by induced gluon bremsstrahlung as illustrated in Figs.~\ref{fig:twist_four}(b)-(d).
The induced gluon bremsstrahlung effectively reduces the energy of the leading quark jet and
leads to a medium modified quark fragmentation function. We will work in collision frame where
the momentum of virtual photon, the momentum of target nucleus per nucleon and the
radiated gluon are,respectively,
\begin{eqnarray}
q=[-Q^{2}/2q^{-},q^{-},\vec 0_{T}], \nonumber \\
p=[p^{+},0,\vec 0_{T}],\nonumber \\
l=[l_{T}^{2}/2zq^{-},zq^{-},\vec l_{T}],
\end{eqnarray}
and the Bjorken variable is $x_{B}=Q^{2}/2q^{-}p^{+}$. In this paper, we will use the convention for
light-cone momentum ($p$) and coordinate ($y$) variables
\begin{eqnarray}
p&=&[p^{+},p^{-},\vec p_{T}], \;\; p^{+}=p_{0}+p_{z}; \;\;\; p^{-}=(p_{0}-p_{z})/2, \\
y&=&[y^{+},y^{-},\vec y_{T}],\;\; y^{+}=t+z;\;\; y^{-}=(t-z)/2.
\end{eqnarray}

Within the framework of a generalized
factorization of higher-twist contribution to the semi-inclusive cross section from multiple parton
scattering, medium modification to the quark fragmentation function was obtained from the
double scattering in Figs.~\ref{fig:twist_four}(b)-(d), the interference between single and triple  parton
scattering and the corresponding virtual corrections \cite{Guo:2000nz,Wang:2001ifa,Zhang:2003yn},
\begin{eqnarray}
\nonumber
\label{eq:modified_D}
 \Delta D_{q}^h(z_h,Q^2)&=& \int_0^{Q^2}\frac{dl_{T}^{2}}{l_{T}^{2}}
 \frac{\alpha_{s}}{2\pi}
\int _{z_{h}}^{1} \frac{dz}{z}
 [\Delta \gamma_{q\rightarrow qg}(z,x_{B},x_L,l_T^2)D_{q}^h(z_h/z,Q^{2})\\
&&+ \Delta \gamma_{q\rightarrow gq}(z,x_{B},x_L,l_T^2)D_{g\rightarrow h}(z_h/z,Q^{2}) ],
\end{eqnarray}
where $D_{q}^h(z,Q^{2})$ and $D_{g\rightarrow h}(z,Q^{2})$ are the fragmentation functions
in vacuum and the modification to the splitting functions,
\begin{eqnarray}
\label{eq:qg & gq}
\nonumber
 \Delta \gamma_{q\rightarrow qg}(z,x_{B},x_L,l_T^2)&=\frac{C_A}{l_T^2}&\left[ \frac{1+z^2}{(1-z)_+}T_{qg}^A(x_{B},x_L)+\delta(1-z)\Delta^{(q)} T_{qg}^A(x_{B},l_T^2)\right]\\
\nonumber
&&\times\frac{2\pi \alpha_s}{N_c f_q^A(x_{B})}\\
 \Delta \gamma_{q\rightarrow gq}(z,x_{B},x_L,l_T^2)&=&\Delta \gamma_{q\rightarrow qg}(1-z,x_{B},x_L,l_T^2)
\end{eqnarray}
are obtained from the induced gluon bremsstrahlung spectra and therefore are related
to the twist-four nuclear quark-gluon correlation distribution,
\begin{eqnarray}
\hspace{-0.2in}T^A_{qg}(x_{B},x_L)&=& \int \frac{dy^{-}}{2\pi}\, dy_1^-dy_2^-
e^{i(x_{B}+x_L)p^+y^-}(1-e^{-ix_Lp^+y_2^-})
(1-e^{-ix_Lp^+(y^--y_1^-)}) \nonumber \\
&&\hspace{-0.4in}\times\frac{1}{2}\langle A | \bar{\psi}_q(0)\,
\gamma^+\, F_{\sigma}^{\ +}(y_{2}^{-})\, F^{+\sigma}(y_1^{-})\,\psi_q(y^{-})
| A\rangle \theta(-y_2^-)\theta(y^- -y_1^-).
\label{eq:qgmatrix}
\end{eqnarray}
The matrix element,
\begin{eqnarray}
\Delta^{(q)} T^A_{qg}(x,\ell_T^2) &\equiv &
\int_0^1 dz\frac{1}{1-z}\left[ 2 T^A_{qg}(x,x_L)|_{z=1}
-(1+z^2) T^A_{qg}(x,x_L)\right] \, , \label{eq:vsplit2}
\end{eqnarray}
in the second term (which is proportional to a $\delta(1-z)$ function) of the modification to the splitting function
comes from virtual corrections to the multiple parton scattering cross section. This term can be
constructed from the momentum sum rule (or momentum conservation) for the modified
fragmentation function \cite{Guo:2000nz,Wang:2001ifa,Zhang:2003yn},
\begin{equation}
\int dz z \Delta D_{q}^h(z,Q^{2})=0.
\end{equation}

The quark distribution function $f_{q}^{A}(x_{B})$ of the nucleus is defined as
\begin{equation}
f^A_{q}(x_{B})= \int \frac{dy^{-}}{2\pi} e^{ix_{B}p^+y^-} \frac{1}{2}\langle A | \bar{\psi}_q(0)\,
\gamma^+\, \psi_q(y^{-})| A\rangle .
\end{equation}

%The twist-four quark-gluon correlation distribution in Eq.~(\ref{eq:qgmatrix}) in fact contains
%four separate twist-four matrix elements corresponding to two different physical processes
%in Figs.~\ref{fig:twist_four}(b)-(d) and  their interferences. When the radiated gluon is induced
%by the secondary scattering, the struck quark carrying momentum fraction $x_{B}$ is on shell
%immediately following its interaction with the virtual photon and the second gluon ( refereed to as hard)
%from nuclear medium should carry finite momentum fraction of the target nucleon,
%\begin{equation}
%x_{L}=\frac{l_{T}^{2}}{2p^{+}q^{-}z(1-z)}.
%\end{equation}
%If the radiated gluon, on the other hand, is produced through the final state radiation of the
%quark-photon interaction, the initial quark carries momentum fraction $x_{B}+x_{L}$ and
%becomes on shell only after the gluon radiation. It subsequently interacts with a soft
%gluon (carrying almost zero momentum fraction). These two different processes will also
%interfere with each other leading to the Landau-Pomeranchuck-Migdal (LPM) \cite{Landau:1953gr,Migdal:1956tc}
%interference which suppresses the spectra of radiative gluons whose formation time
%\begin{equation}
%\tau_{f}=\frac{1}{x_{L}p^{+}}=\frac{2q^{-}z(1-z)}{l_{T}^{2}},
%\end{equation}
%is much larger than the medium length.

With this modification to the fragmentation functions from single induced gluon emission, one can
calculate the modified fragmentation functions by including the vacuum fragmentation functions which
satisfy the vacuum DGLAP evolution equations,
\begin{equation}
\tilde{D}_{q}^h(z,Q^{2})=D_{q}^h(z,Q^{2})+\Delta D_{q}^h(z,Q^{2}).
\end{equation}
If one defines the parton energy loss as the energy carried away by the radiated gluon, one obtains
\begin{eqnarray}
\label{eq:delta_z}
  \frac{\Delta E}{E}&=& \langle\Delta z_g\rangle=\frac{\alpha_s}{2\pi}\int \frac{d l_T^2}{l_T^2}
 \int d z z \Delta \gamma_{q\rightarrow gq}(z,l_T^2) \nonumber \\
&=&\alpha_{s}^{2 }\int d l_T^2\int_0^1 d z \frac{C_{A}}{N_{c}}
\frac{1+(1-z)^2}{l_T^4}\frac{T_{qg}^A(x_B,x_L)}{f_q^A(x_B)}.
\end{eqnarray}

Assuming factorization of the two parton correlation distribution $T_{qg}^{A}$ in terms of quark
and gluon distributions in nucleons \cite{Osborne:2002st,CasalderreySolana:2007sw},
one can express the twist-four nuclear quark-gluon correlation distribution,
\begin{eqnarray}
\frac{2\pi\alpha_s}{N_c}\frac{T_{qg}^A(x_B,x_L)}{f_q^A(x_B)}&=&
\frac{2\pi\alpha_s}{N_c} \pi \int dy^{-} \rho_{N}^{A}(y)[1-\cos(x_{L}p^{+}y^{-})]
\nonumber \\
&&\hspace{1.0in} \times \left [(x_{L}G_{N}(x_{L})+c(x_{L})[xG_{N}(x)]_{x\approx 0}\right] \nonumber \\
&=&\int dy^-[1-\cos(x_{L}p^{+}y^{-})][\hat q_F(x_{L},y)+c(x_{L})\hat q_{F}(0,y)]
\nonumber \\
\end{eqnarray}
in terms of the generalized jet transport parameter in nuclear medium,
\begin{equation}
\hat q_{R}(x_{L},y)=\frac{4\pi^2\alpha_sC_{R}}{N_{c}^{2}-1}\rho_{N}^{A}(y)x_{L}G_{N}(x_{L}),
\end{equation}
which in the limit of $x_{L}=0$ is also the transverse momentum broadening per unit distance for a parton in the
$R$-representation of color. Here $c(x_{L})=f_{q}^{A}(x_{B}+x_{L})/f_{q}^{A}(x_{B})$. The momentum
fraction $x_L$ dependence in the generalized transport parameter reflects the
energy transfer to the medium parton during the inelastic scattering with the propagating parton. Such recoil of
the medium parton is in fact the cause for elastic energy loss \cite{Wang:2006qr}. We will limit the study to gluon
bremsstrahlung and therefore will not consider the $x_{L}$ dependence of the generalized jet transport
parameter $\hat q(x_{L},y)\approx \hat q(0,y)\equiv \hat q(y)$. Under this approximation,
\begin{equation}
\frac{2\pi\alpha_s}{N_c}\frac{T_{qg}^A(x_B,x_L)}{f_q^A(x_B)} \approx
 \int dy^- \hat q_{F}(y) 4 \sin^{2}(x_{L}p^{+}y^{-}/2) \, .
\end{equation}

Since $x_L=\infty$ for $z=0$ and both the quark $f_q^A(x_B+x_L)$ and gluon distributions $x_LG_N(x_L)$
vanish at $x_L=\infty$, the matrix elements related to the virtual correction in the modified fragmentation function
becomes
\begin{equation}
\frac{2\pi\alpha_s}{N_c} \frac{\Delta^{(q)} T^A_{qg}(x,\ell_T^2)}{f_q^A(x_B)} \approx
- \int_{\epsilon}^{1-\epsilon} dz\frac{1+z^2}{1-z} \int dy^{-} \hat q_{F}(y) 4 \sin^{2}(x_{L}p^{+}y^{-}/2) \, , \label{eq:vsplit}
\end{equation}
where
\begin{equation}
\epsilon(l_{T}^{2})=\frac{1}{2}\left(1-\sqrt{1-2l_T^2/p^{+}q^{-}}\right); \,\, l_{T}^{2}\leq 2 p^{+}q^{-}
\label{eq:cut}
\end{equation}
are the restrictions imposed on the limits of $z$ and $l_{T}^{2}$ integrations by requiring $x_{L}\le 1$. Such limits
are also imposed in the calculation of the modified fragmentation function. The radiated parton energy
loss from single secondary scattering is then
\begin{eqnarray}
\label{eq:eloss}
  \frac{\Delta E}{E} &=&C_{A} \frac{\alpha_{s}}{2\pi} \int dy^{-} \int_{0}^{Q^{2}} \frac{d l_T^2}{l_{T}^{4}}
  \int_\epsilon^{1-\epsilon} d z
[1+(1-z)^2] \hat q_{F}(y) 4 \sin^{2}(x_{L}p^{+}y^{-}/2).
\end{eqnarray}

\section{Modified DGLAP evolution equations}

To take into account of multiple induced gluon emissions, one can follow the resummation of
gluon bremsstrahlung in vacuum and assume that multiple medium induced bremsstrahlung
can be resummed in the same way and one can obtain the mDGLAP evolution equations
for the modified fragmentation functions \cite{Guo:2000nz,Wang:2001ifa},
\begin{eqnarray}
 \label{eq: modified DGLAP1}
\nonumber
 \frac{\partial \tilde{D}_q^h(z_h,\mu^2)}{\partial \ln \mu^2}&=&\frac{\alpha_s(\mu^2)}{2\pi}\int_{z_h}^1
 \frac{dz}{z}\left [ \tilde{\gamma}_{q\rightarrow qg}(z,\mu^2)\tilde{D}_q^h(\frac{z_h}{z},\mu^2)\right. \\
&& \hspace{1.5in} \left. +\tilde{\gamma}_{q\rightarrow gq}(z,\mu^2)\tilde{D}_g^h(\frac{z_h}{z},\mu^2)\right ] ,\\
\nonumber
 \label{eq: modified DGLAP2}
  \frac{\partial \tilde{D}_g^h(z_h,\mu^2)}{\partial\ln \mu^2}&=&\frac{\alpha_s(\mu^2)}{2\pi}\int_{z_h}^1
 \frac{dz}{z}\left [ \sum_{q=1}^{2n_f}\tilde{\gamma}_{g\rightarrow q\bar q}(z,\mu^2)\tilde{D}_q^h(\frac{z_h}{z},\mu^2) \right. \\
&& \hspace{1.5in} +\left. \tilde{\gamma}_{g\rightarrow gg}(z,\mu^2)\tilde{D}_g^h(\frac{z_h}{z},\mu^2)\right ] ,
\end{eqnarray}
where the modified splitting functions are given by the sum of the vacuum ones $\gamma_{a\rightarrow bc}(z)$
and the medium modification $\Delta \gamma_{a\rightarrow bc}(z,l_{T}^{2})$,
\begin{equation}
\tilde \gamma _{a\rightarrow bc}(z,l_{T}^{2})=\gamma_{a\rightarrow bc}(z)
    + \Delta \gamma_{a\rightarrow bc}(z,l_{T}^{2})
\end{equation}
with $ \Delta \gamma_{a\rightarrow bc}(z,l_{T}^{2})$
for $q\rightarrow qg$  given in Eq.~(\ref{eq:qg & gq}). In the above mDGLAP evolution equations we
deliberately used $\mu^{2}$ to denote the evolving scale whose maximum value in DIS is $Q^{2}$

In addition to the first mDGALP evolution equation for the modified quark fragmentation function
in Eq.~(\ref{eq: modified DGLAP1}) as derived in Refs.~\cite{Guo:2000nz,Wang:2001ifa} in DIS
off nuclei, one has to consider multiple scattering and induced gluon bremsstrahlung for a gluon jet in
order to complete the mDGLAP evolution equation for medium modified gluon fragmentation
function in Eq.~(\ref{eq: modified DGLAP2}) and the corresponding modified splitting functions.
In the study of contributions from quark-quark (antiquark) double scattering to the modified quark
fragmentation functions within the same framework of generalized factorization
of multiple parton scattering \cite{Schafer:2007xh}, one can in fact relate the effective
splitting functions to the corresponding parton-parton scattering amplitudes. From gluon-gluon
scattering matrix elements, one can obtain the following medium modification to the splitting
functions (see Appendix 3 in Ref.~\cite{Schafer:2007xh}) due to double scattering between
a gluon jet and medium gluons,
\begin{eqnarray}
\Delta \gamma_{g\rightarrow q\bar q}(z,l_{T}^{2})
%%&=&\frac{1}{2 l_{T}^{2}}
%%[z^{2}+(1-z)^{2}]\left[1-\frac{N_{c}}{C_{F}}z(1-z)\right]\frac{2\pi\alpha_{s}}{N_{c}}
%%\frac{T_{gg}^{A}}{f_{g}^{A}} \nonumber \\
&=&\frac{1}{2 l_{T}^{2}} [z^{2}+(1-z)^{2}]\left[1-\frac{N_{c}}{C_{F}}z(1-z)\right] \nonumber \\
&& \hspace{1.in} \times \int dy^{-} \hat q_{F}(y) 4 \sin^{2}(x_{L}p^{+}y^{-}/2) \\
\Delta \gamma_{g\rightarrow gg }(z,l_{T}^{2})
%%&=&\frac{2C_{A}}{l_{T}^{2}C_{F}}
%%\frac{(1-z+z^{2})^{3}}{z(1-z)}2\pi\alpha_{s} \frac{T_{gg}^{A}}{f_{g}^{A}} \nonumber \\
&=&\frac{2C_{A}N_c}{l_{T}^{2}C_{F}}\frac{(1-z+z^{2})^{3}}{z(1-z)_{+}}
\int dy^{-} \hat q_{F}(y) 4 \sin^{2}(x_{L}p^{+}y^{-}/2) \nonumber \\
&& -\delta(1-z)\frac{\Delta_{g}(l_{T}^{2})}{l_{T}^{2}} \, ;\\
\Delta_{g}(l_{T}^{2})&=&\int dy^{-} \int_{\epsilon}^{1-\epsilon} dz \left\{
n_f [z^{2}+(1-z)^{2}]\left[1-\frac{N_{c}}{C_{F}}z(1-z)\right] \right. \nonumber \\
&+& \left. \frac{2C_AN_c}{C_F}\frac{(1-z+z^{2})^{3}}{z(1-z)} \right\}
  \hat q_{F}(y) 4 \sin^{2}(x_{L}p^{+}y^{-}/2) \, ,
 \end{eqnarray}
where we have similarly assumed that the parton correlation distribution vanishes for $x_{L}\ge 1$
which imposes limit on the integration over $\epsilon (l_{T}^{2}) \le z \le 1-\epsilon(l_{T}^{2})$.
The above calculations of the modification to the splitting functions are more complete since they
include non-leading terms when $1-z\rightarrow 0$. For mDGLAP evolution equation for the
quark fragmentation function, one can also compute the modification to the splitting function
for $q\rightarrow qg$ from the complete matrix element of quark-gluon
Compton scattering ((see Appendix 3 in Ref.~\cite{Schafer:2007xh}),
\begin{eqnarray}
\Delta \gamma_{q\rightarrow qg}(z,l_{T}^{2})
%%&=&\frac{1}{l_{T}^{2}}
%%\left[ C_{A}\frac{z(1+z^{2})}{1-z} + C_{F}(1-z)(1+z^{2})\right]\frac{2\pi\alpha_{s}}{N_{c}}
%%\frac{T_{qg}^{A}}{f_{q}^{A}} \nonumber \\
&=&\frac{1}{l_{T}^{2}} \left[ C_{A}\frac{z(1+z^{2})}{(1-z)_{+}} + C_{F}(1-z)(1+z^{2})\right]
\nonumber \\
&&\hspace{0.3in} \times \int dy^{-} \hat q_{F}(y) 4 \sin^{2}(x_{L}p^{+}y^{-}/2)
-\delta(1-z)\frac{\Delta_{q}(l_{T}^{2})}{l_{T}^{2}};\\
\Delta_{q}(l_{T}^{2})&=&\int dy^{-} \int_{\epsilon}^{1-\epsilon} dz
\left[ C_{A}\frac{z(1+z^{2})}{1-z} + C_{F}(1-z)(1+z^{2})\right] \nonumber \\
&&\hspace{1.in} \times  \hat q_{F}(y) 4 \sin^{2}(x_{L}p^{+}y^{-}/2) \, .
  \end{eqnarray}

Note that we have expressed the medium modifications to the splitting functions in mDGLAP
evolution equations for both quark and gluon fragmentation functions in terms
of the quark jet transport parameter $\hat q_{F}$ and leave the color factors explicitly in
the $z$-dependent parts. There is no single overall color factor for each effective splitting
functions because they involve more than one channel of quark-gluon or gluon-gluon scattering
which have different color factors. However, in the limit of soft gluon radiation, $z\rightarrow 1$
(here $z$ is the momentum fraction carried by the leading parton), the modifications are dominated
by leading term $1/(1-z)$. In this case,
\begin{equation}
\Delta \gamma_{g\rightarrow gg }(z,l_{T}^{2})\approx \frac{N_{c}}{C_{F}}\Delta \gamma_{q\rightarrow qg }(z,l_{T}^{2}),
\,\,\, {\rm for}\,\, z\rightarrow 1,
\label{eq:color}
\end{equation}
and $N_{c}/C_{F}=2N_{c}^{2}/(N_{c}^{2}-1)=9/4$.

We can also similarly consider the quark-quark (antiquark) scattering and quark-antiquark annihilation processes
for the secondary scattering. These processes will be responsible for flavor changing (conversion) in the
parton propagation. They generally involve quark or antiquark density distributions of the medium and
are also suppressed by a factor of $l_{T}^{2}/Q^{2}$ ~\cite{Schafer:2007xh}. We will neglect these
processes for now and focus on the most dominant gluon radiation in the mDGLAP equations.

In the following we will numerically solve the mDGLAP evolution equations for the modified fragmentation
functions using a modified HOPPET (Higher Order Perturbative Parton Evolution Toolkit) \cite{hoppet} in LO.
HOPPET is a Fortran95 package with the GNU Public License that carries out the vacuum DGLAP
evolution in $z$-space using Runge-Kutta method with a given initial condition $D_{a}(z,Q_0^2)$.
We replace the normal LO splitting functions in HOPPET with the modified splitting functions
in each step of Runge-Kutta iteration to solve the mDGLAP evolution for modified fragmentation functions.

Since all the medium modified splitting functions in mDGLAP evolution equations are proportional to the path
integral of the jet transport parameter $\hat q_{F}(y)$ along the parton propagation length, the modified
fragmentation functions will be determined completely by the spatial profile (and time evolution in the case
of heavy-ion collisions) of the $\hat q_{F}(y)$ which characterizes the properties of the medium that
a jet probes. Therefore, we will study in the rest of this paper the dependence of the modified fragmentation
functions on the profile of the jet transport parameter, its value, length of the medium, in addition to
the energy ($E$) and scale ($Q^{2}$) dependence. By setting $\hat q_{F}(y)=0$, the mDGLAP evolution
equations will become those in vacuum and the corresponding vacuum fragmentation functions and their
$Q^{2}$ dependence will be recovered.

 \section{The ``brick'' problem}

We will consider first the simplest profile of the medium characterized by a constant value
of $\hat q_{F}=\hat q_{0}$ for a uniform medium with finite length $L$, assuming that a jet
is produced at $y=0$ via DIS and propagates through the medium and finally hadronizes outside
the medium. Such a ``brick'' problem is illustrative for understanding the main features of the
modified fragmentation function from mDGLAP evolution equations. When solving mDGLAP
evolution equations in this paper, we will use their LO form with $n_{f}=3$ number of quark
flavors and a running strong coupling constant $\alpha_{s}(Q^{2})$ with $\Lambda_{\rm QCD}=214$ MeV.

\subsection{Initial conditions}
\label{sub-initial}

To solve the mDGLAP evolution equations one also has to specify the fragmentation functions at an
initial scale $Q_{0}^{2}$ as the initial condition. Such initial conditions in principle should be different
in medium and vacuum. Since perturbative QCD cannot be applied below the initial scale $Q_{0}^{2}$
the initial conditions are normally supplied by experimental
measurements which are currently not available for medium modified fragmentation functions.
Without such information, we will have to resort to a model of the medium modified fragmentation
function at the initial scale $Q_{0}^{2}$. Many studies \cite{Armesto:2007dt,Majumder:2009zu}
with the modified DGLAP evolution approach have
simply assumed the initial condition at $Q_{0}^{2}$ in medium as the same in vacuum. This implies that
partons below scale $Q_{0}^{2}$ will not interact with the medium and therefore suffer no energy loss.
This is apparently counterintuitive in theory. Phenomenologically, such an assumption also means that
all medium modification comes from processes above scale $Q_{0}^{2}$ and therefore has a very strong
$Q^{2}$ dependence, which contradicts the experimental data as we will see later. To take into account
medium modification to the fragmentation functions below the initial scale $Q_{0}^{2}$, we will assume in
this study,
\begin{equation}
 \tilde D_{a}(Q_0^2)=D_{a}(Q_0^2)+\Delta D_{a}(Q_{0}^{2})\, , \,\,\, a=g,q,\bar q.
 \end{equation}
where $D_{a}(Q_0^2)$ is the vacuum fragmentation function and $\Delta D_{a}(Q_{0}^{2})$ will be
generated purely from medium via the mDGLAP equations (\ref{eq: modified DGLAP1}) starting
at $\mu^{2}=0$. In practice, we will simply set the effective splitting functions to contain only the
medium-induced parts,  $\tilde\gamma_{a\rightarrow bc}(z,l_{T}^{2})=\Delta\gamma_{a\rightarrow bc}(z,l_{T}^{2})$,
freeze the running strong coupling constant at $\mu^{2}=Q_{0}^{2}$ and use the vacuum fragmentation functions
as initial condition at $\mu^{2}=0$ in the calculation of $\Delta D_{a}(Q_{0}^{2})$. The value of $Q_{0}^{2}$ will
be set at $Q_{0}^{2}=1$ GeV$^{2}$ in this study unless otherwise stated.

\subsection{Momentum sum rule}

To test the modified HOPPET method and estimate errors of the numerical solution to the mDGLAP evolution
equations,  we first check the following momentum sum rules
\begin{equation}
\sum_h\int_0^1\mathrm{d}z z\tilde D_a^h(z,Q^2)=1,
\end{equation}
which should be satisfied for all parton fragmentation functions at any scale $Q^2$. For this test, we
assume the following initial fragmentation functions,
\begin{equation}
\label{eq:delta_initial_all}
D_{q}^h(z,Q_0^2)=D_{\bar{q}}^h(z,Q_0^2)=D_{g}^h(z,Q_0^2)=\delta(1-z).
\end{equation}
which corresponds to a parton-hadron duality in vacuum at scale $Q_{0}^{2}$ with just one hadron species.
Medium modification of the initial condition at $Q_{0}^{2}$ will be given according to the
prescription in Sec.~\ref{sub-initial}.

\begin{figure}
  \centering
  \includegraphics[width=0.6\textwidth]{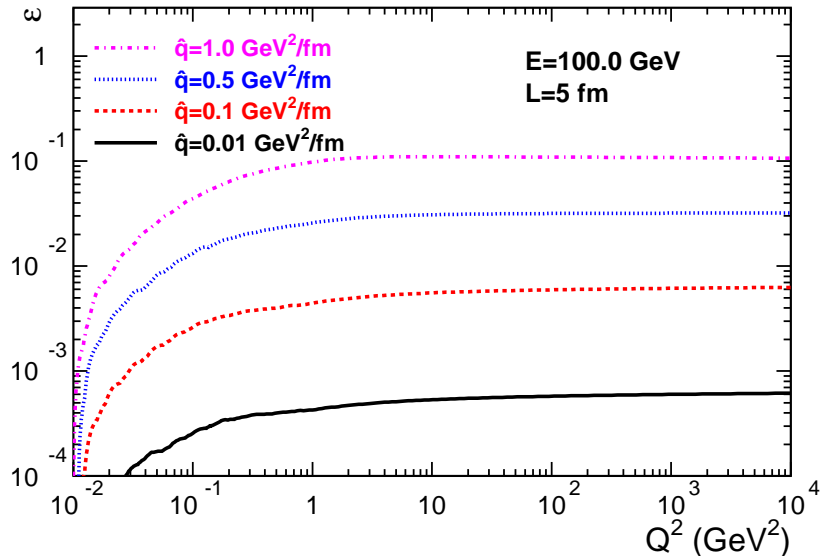}%
  \caption{(color online) Numerical violation of the momentum sum rule for the fragmentation function of a $u$ quark as
  a function of the momentum scale $Q^{2}$ and for different values of the jet transport parameter $\hat q_{0}$.}
  \label{fig:sum_rule}
\end{figure}

Shown in Fig.~\ref{fig:sum_rule}, is the numerical violation of the above momentum sum rule for $u$ quark
fragmentation functions
\begin{equation}
\varepsilon=1-\int_0^1 d z z \tilde D_u^h(z,Q^2),
\end{equation}
in medium with different values of jet transport parameter $\hat q_{0}$.
The sum rule for the fragmentation function in vacuum ($\hat{q}_0=0$) is almost
perfect and is also the case for small values of jet transport parameter $\hat q_{0}\leq 0.1$ GeV$^{2}$/fm.
The numerical error becomes larger as the value of $\hat{q_0}$ is increased in the
``brick'' medium. The HOPPET method uses a grid in $z$-space to numerically evaluate integration in $z$.
This grid in $z$ has a minimum value $z_{min}$.  Therefore the numerical integration over $z$ is always
truncated in $z$ space below $z_{min}$. Even though $z_{min}=e^{-30}$ in this study is a very small
number but not exactly zero.
So, the momentum fraction contained in the $z<z_{min}$ region causes the numerical violation of the momentum
sum rule. Such violation should increase with $Q^{2}$ and $\hat q_{0}$ because the mDGLAP evolution
puts more hadrons (or partons) in the small $z<z_{min}$ region at larger $Q^{2}$ or for larger values of $\hat q_{0}$.
However, the sum rules in medium are still satisfied with an error better than 10 percent for most of the range
of $Q^{2}$ and $\hat q_{0}$ as shown in Fig.~\ref{fig:sum_rule}.

\subsection{Shower parton distributions and energy loss}

Before we turn our attention to the medium modified fragmentation functions, it is useful to study the
shower parton distributions generated by a propagating parton which can provide useful information
about the underlying parton production and energy loss before the final hadronization outside the medium. To this end, we
again assume parton-hadron duality for three species of ``hadrons'', $h=q,\bar q, g$ and
take initial conditions,
\begin{eqnarray}
\label{eq:delta_initial_u}
\nonumber
D_{a}^a(z,Q_0^2)&=&\delta(1-z); D_{a}^{b\neq a} (z,Q_0^2)=0\,\,\, (a,b=q,\bar q, g),
\end{eqnarray}
which simply mean that partons stop branching in vacuum at scale $Q_{0}^{2}$. A parton with a large
scale $Q^{2}>Q_{0}^{2}$, however, will undergo both vacuum and medium induced gluon bremsstrahlung and
pair production as described by the mDGLAP evolution equations in Eqs.~(\ref{eq: modified DGLAP1})
and (\ref{eq: modified DGLAP2}). The original parton will then generate other species of partons within
its parton shower. The initial condition for medium modification will be given by the prescription
in Sec. \ref{sub-initial}, which implies that jet-medium interaction continues to generate parton shower
below scale $Q_{0}^{2}$.

\begin{figure}
  \centering
     \includegraphics[width=\textwidth]{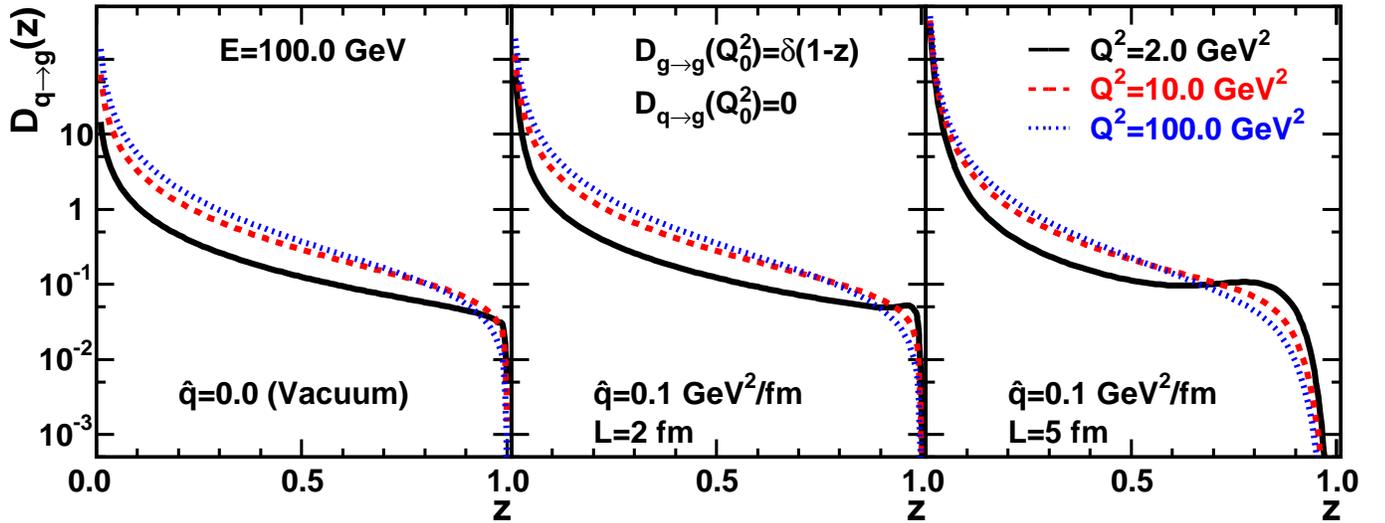}
   \caption{(color online) Gluon distribution in the parton shower of a quark jet}
  \label{fig:gluon}
\end{figure}

\begin{figure}
  \centering
 \includegraphics[width=\textwidth]{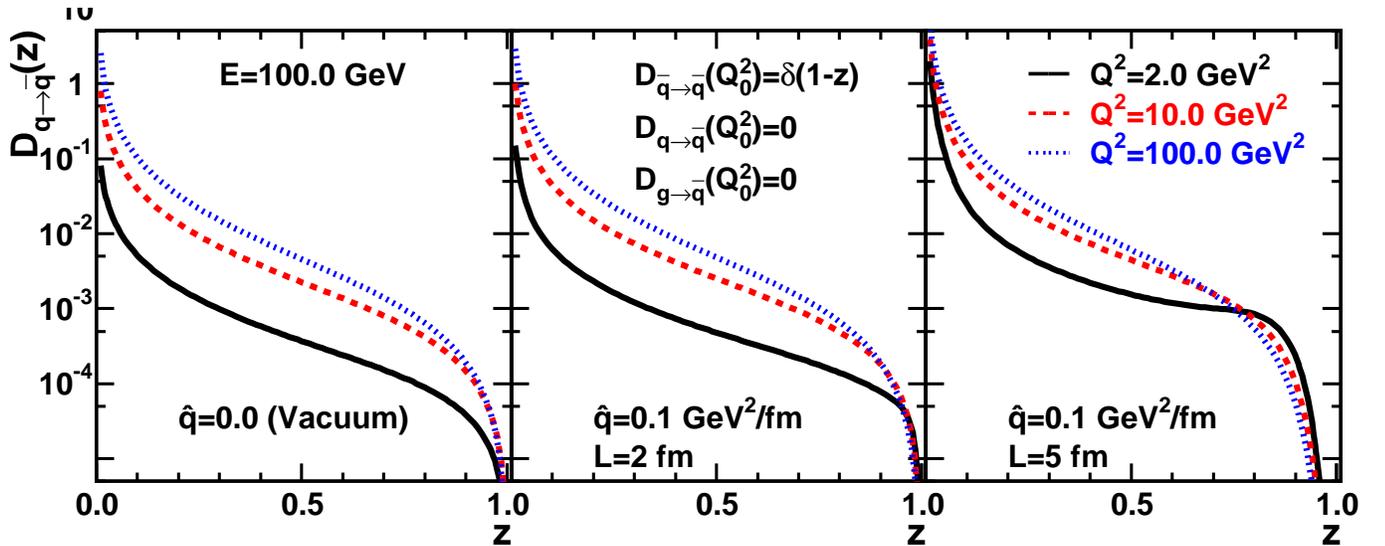}
   \caption{(color online) Sea quark distribution in the parton shower of a quark jet}
  \label{fig:sea}
\end{figure}

Shown in Figs.~ \ref{fig:gluon} and \ref{fig:sea} are gluon  $\tilde D_{q}^{g}(z,Q^{2})$  and
antiquark (or sea quark) distribution $\tilde D_{q}^{\bar q}(z,Q^{2}) $ within a quark jet for different values
of jet transport parameter $\hat q_{0}$, medium
length $L$ and momentum scale $Q^{2}$.  We have set $Q_{0}^{2}=1$ GeV$^{2}$ and initial parton
energy $E=100$ GeV. Yields of gluons and sea quarks within a quark
jet increase with $Q^{2}$  for both vacuum (left panels) and medium
modified (middle and right panels) fragmentation functions and the spectra also soften as more
gluon bremsstrahlung and pair production take place during the evolution with $Q^{2}$. For small $Q^{2}$
the modified shower parton distributions become harder at large $z$ due to medium-induced pair
production of quark-anti-quark and gluons (in the $g\rightarrow gg$ splitting). However, with increasing
$Q^{2}$ the modified fragmentation functions again soften and the yields increase with $\hat q_{0}$ and $L$
due to the additional gluon bremsstrahlung and pair production induced by jet quenching during the propagation of the
quark jet in medium. The sea quark distributions are generally softer than gluon distributions within a
quark jet because quark-antiquark pairs are produced through gluon fission in vacuum and fusion in
medium.

\begin{figure}
  \centering
\includegraphics[width=\textwidth]{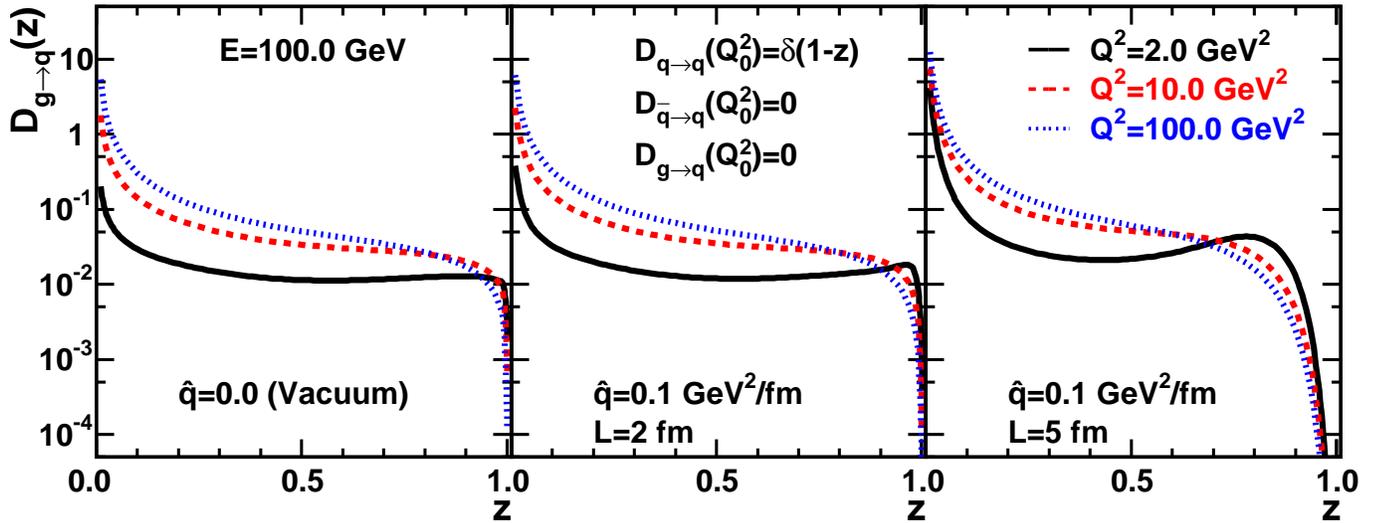}
   \caption{(color online) Quark (antiquark) distribution in the parton shower of a gluon jet}
  \label{fig:gq}
\end{figure}

\begin{figure}
  \centering
  \includegraphics[width=\textwidth]{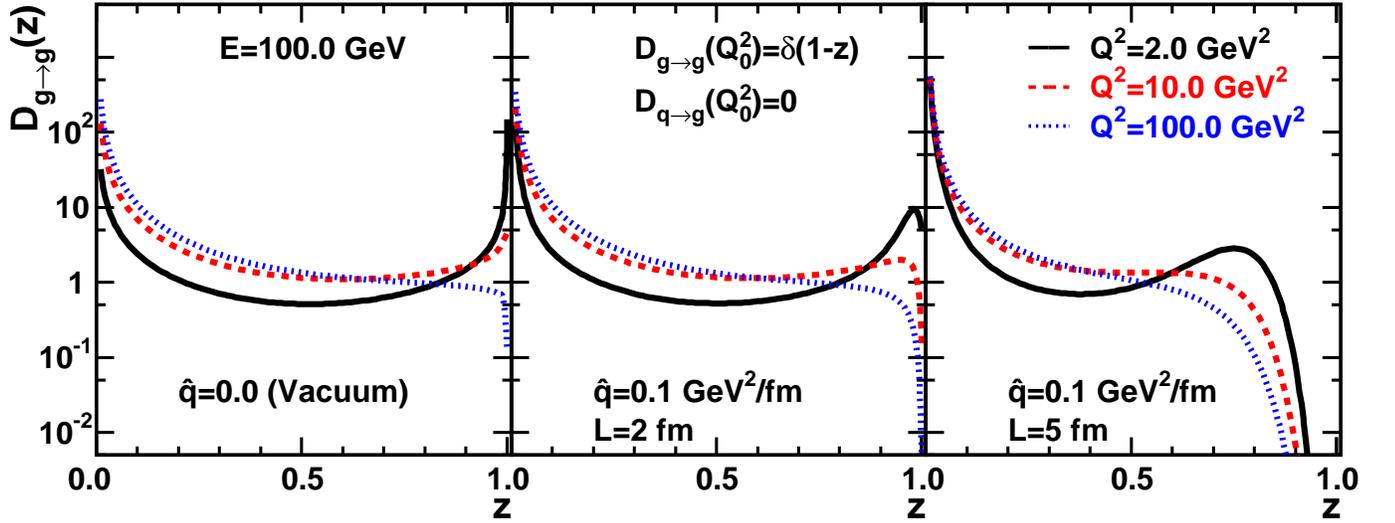}
   \caption{(color online) Gluon distribution in the parton shower of a gluon jet.}
  \label{fig:gg}
\end{figure}

Similar behavior is also seen in quark(antiquark) distribution $\tilde D_{g}^{q}(z,Q^{2})$ within a gluon jet as
shown in Fig.~\ref{fig:gq}. For small $Q^{2}=2$ GeV$^2$, one notes that the distribution is quite flat,
reflecting the shape of the splitting function $\tilde \gamma_{g\rightarrow q\bar q} (z)$
when the distribution is dominated by pair creation from the leading gluon whose initial distribution is
peaked at $z=1$. As one increases $\hat q_{0}$, $Q^{2}$ and $L$, soft gluons are also generated
which in turn produce soft quark-antiquark pairs. As a consequence, the final quark distribution within
a gluon jet will become softer as the gluon distribution $\tilde D_{q}^{g}(z,Q^{2})$ and sea quark
distribution $\tilde D_{q}^{\bar q}(z,Q^{2})$ within a quark jet in Figs. ~\ref{fig:gluon} and \ref{fig:sea}.
The gluon distribution within a gluon jet on another hand contains both the initial gluon and produced
gluons from vacuum and medium induced bremsstrahlung. As one can see from Fig.~\ref{fig:gg}, it
indeed develops a peak at $z\sim 0$ due to vacuum and induced bremsstrahlung in addition to the
peak at $z=1$ of its initial distribution as one increases the values of $\hat q_{0}$, $L$ and $Q^{2}$.
Since one cannot separate initial and produced gluons, the concept of parton energy loss becomes
ambiguous for a gluon jet in this picture of successive bremsstrahlung in medium.

\begin{figure}
  \centering
   \includegraphics[width=\textwidth]{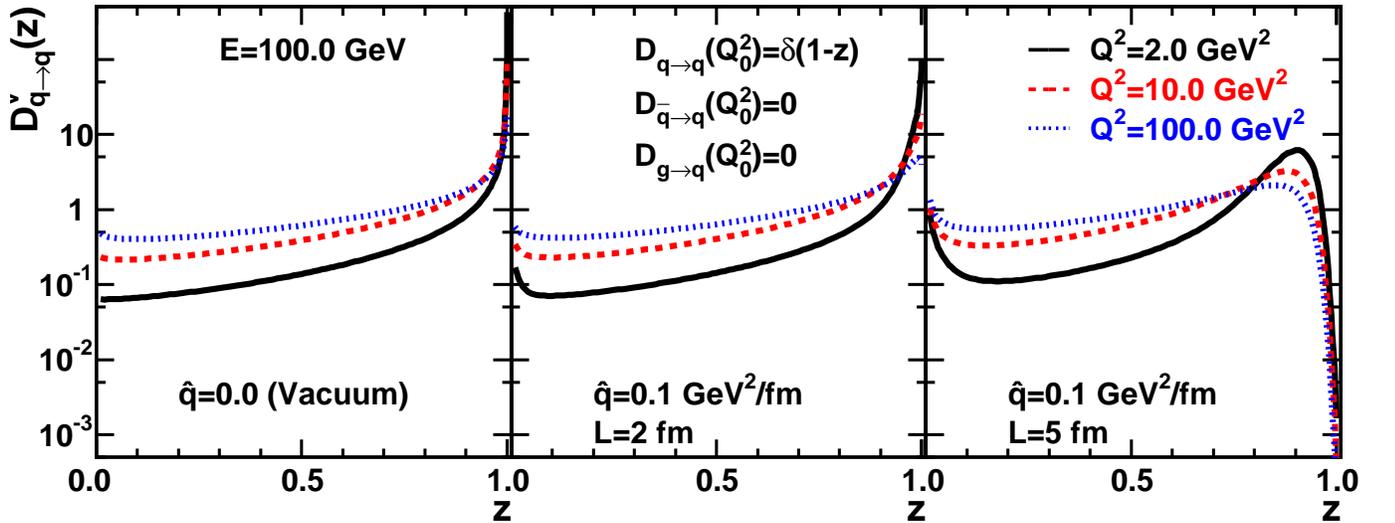}
   \caption{(color online) Valence quark distribution in the parton shower of a quark jet}
  \label{fig:valence}
\end{figure}

The case for a quark jet is different because net quark number is a conserved quantity. Even though
the quark distribution contains both the initial leading or valence quark and the sea quarks from pair
production, the sea quark distribution should be the same as the antiquark distribution and therefore
can be subtracted. For the purpose of studying attenuation of the leading or valence parton
due to vacuum and medium induced gluon bremsstrahlung we define the valence quark distribution as
\begin{equation}
\tilde D_{q}^{v}(z,Q^{2})\equiv \tilde D_{q}^{q}(z,Q^{2})-\tilde D_{q}^{\bar q}(z,Q^{2}).
\end{equation}
Shown in Fig.~\ref{fig:valence} is the valence quark distribution from the mDGLAP evolution equations.
Because of gluon bremsstrahlung and pair production during its evolution and propagation,
the leading or valence quark distribution in both vacuum (left panel) and medium
 (middle and right panel) gradually softens from its initial $\delta (1-z)$
form as one increases the momentum scale $Q^{2}$. Like other fragmentation functions,
significant change of the valence quark distribution occurs already due to vacuum gluon bremsstrahlung
and pair production when it evolves from the initial $Q_{0}^{2}$. As one increases jet quenching
parameter $\hat q_{0}$ and the medium length $L$, the valence quark distribution further softens.

Such softening of the valence quark distribution can be characterized by the total fractional momentum
change afforded by the valence quark for both vacuum and medium modified fragmentation functions
\begin{eqnarray}
 \frac{\Delta E_v}{E}&=&1-\int_0^1 dz\, z  D_q^v(z,Q^{2}); \\
  \frac{\Delta E_m}{E}&=&1-\int_0^1 dz\, z \tilde D_q^v(z,Q^{2}).
  \end{eqnarray}
  The net energy loss due to medium induced gluon bremsstrahlung and pair production is then
  \begin{equation}
  \frac{\Delta E}{E}=\frac{\Delta E_m-\Delta E_{v}}{E}
  =\int_0^1 dz\, z  \left[ D_q^v(z,Q^{2})-\tilde D_q^v(z,Q^{2}) \right].
  \label{eq:eloss-v}
    \end{equation}

\begin{figure}
  \centering
  \includegraphics[width=\textwidth]{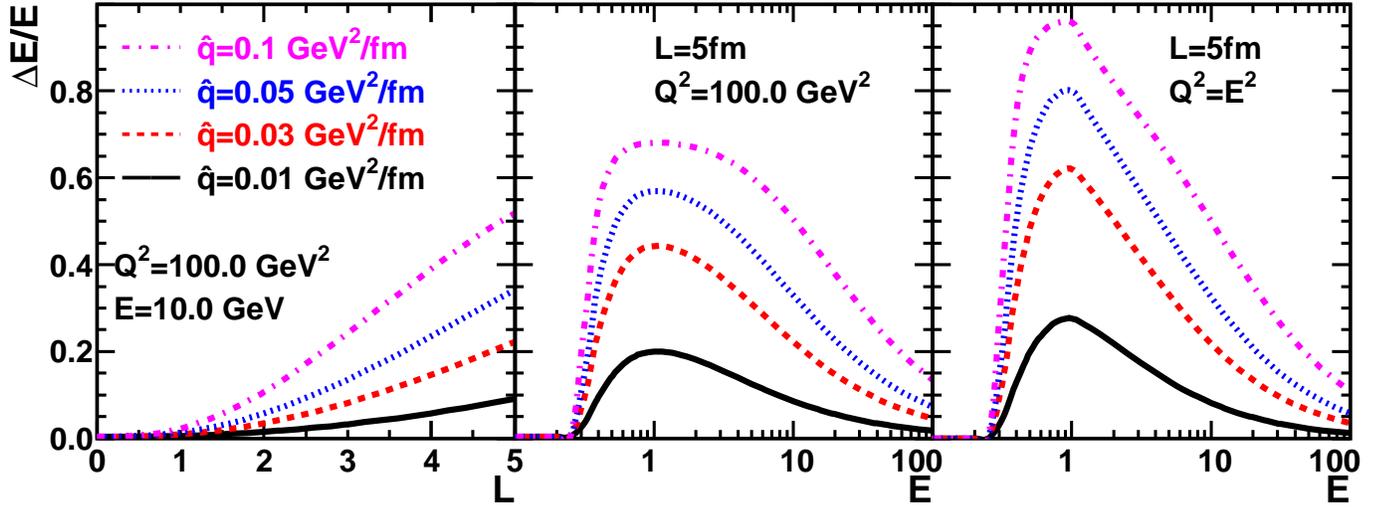}
  \caption{(color online) Energy and length dependence of medium induced energy loss of a valence quark from mDGLAP evolution.}
  \label{fig:loss1}
\end{figure}

\begin{figure}
  \centering
  \includegraphics[width=\textwidth]{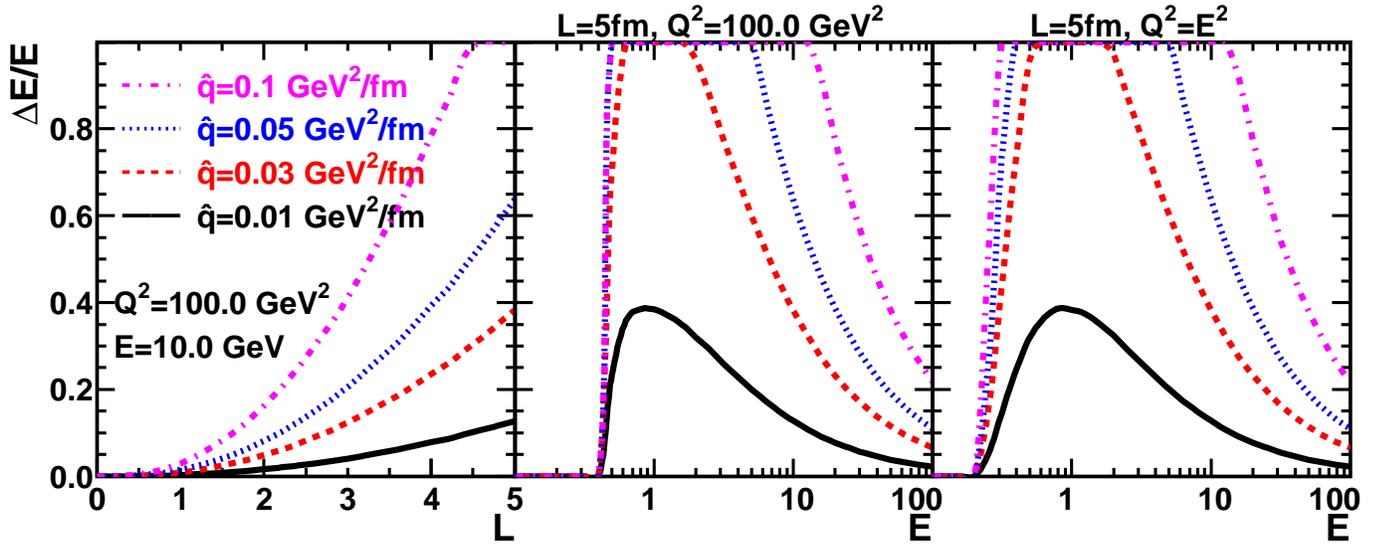}
  \caption{(color online) Energy and length dependence of radiated energy loss of a quark from single gluon bremsstrahlung
  induced by multiple scattering.}
  \label{fig:loss2}
\end{figure}

Shown in Fig.~(\ref{fig:loss1}) is the medium induced energy loss as a function of the medium
length $L$  and initial jet energy $E$ for different values of $\hat q_{0}$. Because of non-Abelian LPM
interference, parton energy loss induced by multiple scattering as seen in the left panel clearly
shows a quadratic dependence on the medium length \cite{Baier:1996sk}. However, for fixed initial
parton energy $E$ and jet transport parameter $\hat q_{0}$, such quadratic length dependence gives
away to a linear dependence for large values of medium length and the induced energy loss even
saturates as $\Delta E/E$ approaches to 1 because energy conservation imposed on each step of
the mDGLAP evolution.

The behavior of such parton energy loss as defined by Eq.~(\ref{eq:eloss-v}) with the modified fragmentation
functions from mDGLAP evolution is quantitatively and sometime qualitatively different from  the averaged
radiative energy loss via a single gluon emission induced by multiple scattering. The averaged energy loss
via single gluon emission is usually identified with the energy carried by the radiated gluon [Eq.~(\ref{eq:eloss})]
as shown in Fig.~\ref{fig:loss2}. Even though energy and
momentum is conserved in the gluon bremsstrahlung process as described by the higher-twist
formalism, energy loss in this case of single gluon emission involves energy loss per emission
multiplied by the average number of scatterings which could become a large number. Such a
calculation apparently becomes invalid when $\Delta E/E\geq 1$. This is an indication of the need
for resummation in terms of mDGLAP evolution in which unitarity of multiple emission is ensured.

In the calculation of
energy loss in both the single gluon emission and the evolution of mDGLAP equations, we have
imposed kinematic cut-off [Eq.~(\ref{eq:cut})] for the integration over fractional longitudinal and transverse
momentum. This is the reason for the strong energy dependence of the energy loss for small values
of $E$. The asymptotic energy dependence of the energy loss $\Delta E$ in both cases is weaker than
linear though the results from mDGLAP evolution shows weaker $E$ dependence than the single gluon
emission.

\subsection{Modified Fragmentation Functions}

In the following, we will solve  mDGLAP evolution equations with the initial conditions $D_{a}^{h}(z,Q_{0}^{2})$
given by the HKN \cite{HKN} parameterization of experimental measurements of fragmentation functions
at $Q_0^2=1$ $GeV^2$ and obtain the corresponding medium modified fragmentation functions for
hadron production. To quantify the medium modification
of the fragmentation functions, we define the modification factor
\begin{equation}
R_{a}^{h}(z,Q^{2})=\frac{\tilde D_{a}^{h}(z,Q^{2})}{ D_{a}^{h}(z,Q^{2})},
\end{equation}
as the ratio between modified and vacuum fragmentation functions.

\begin{figure}
  \centering
 \includegraphics[width=\textwidth]{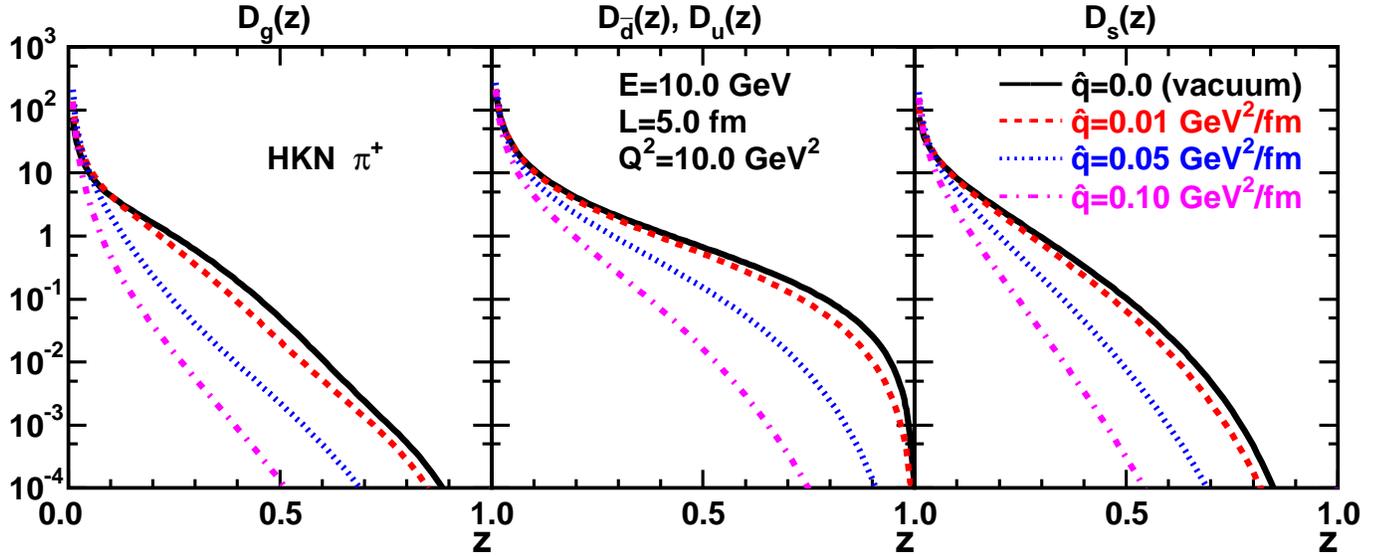}
  \caption{(color online) Modified $\pi$ fragmentation function of a gluon (right), up, down (middle) and strange (left)
  quark jet that is produced at momentum scale $Q^{2}=10$ GeV$^{2}$ with initial energy $E=10$ GeV
  and propagates through a uniform medium with length $L=5$ fm and jet transport parameter $\hat q_{0}=0$
  (vacuum fragmentation function) (solid), 0.01 (dashed), 0.05 (dotted) and 0.1 GeV$^{2}$/fm (dot-dashed). The
  initial condition to the mDGLAP evolution at $Q_{0}^{2}$ is given by HKN \cite{HKN} parameterization .}
  \label{fig:ffqhat}
\end{figure}

\begin{figure}
  \centering
 \includegraphics[width=\textwidth]{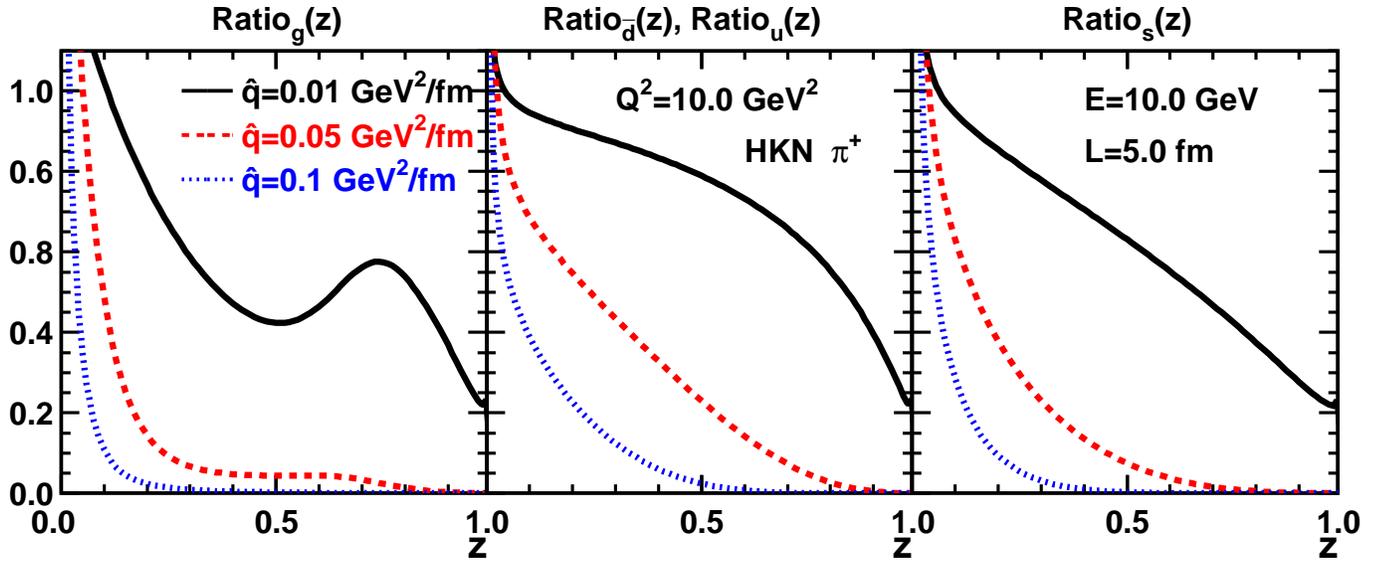}
  \caption{(color online) The modification factor $R_{a}^{\pi}(z,Q^{2})$ for the medium modified fragmentation
  function in Fig.~\ref{fig:ffqhat}.}
  \label{fig:rqhat}
\end{figure}

Shown in Figs.~\ref{fig:ffqhat} and \ref{fig:rqhat} are the modified $\pi^+$ fragmentation functions
for gluon (left panel), $u$, $\bar{d}$ (middle panel) and $s$ (left panel) quark jets and the corresponding
modification factors, respectively, for different values of jet transport parameter $\hat q_{0}$ in a
uniform medium with length $L=5$ fm. The initial jet energy is set to $E=10$ GeV with virtuality
$Q^{2}=10$ GeV$^{2}$. The modified fragmentation functions are seen to be suppressed at medium
and large momentum fraction $z$  with increasing values of $\hat q_{0}$ because of increased
induced gluon bremsstrahlung and pair production. The produced soft gluons and sea quarks will
also hadronize and contribute to the hadron spectra, leading to enhancement of medium modified
fragmentation functions at small $z$. Since the medium modified gluon splitting function is
about 9/4 of a quark [Eq.~(\ref{eq:color})] at large $z$, it radiates more soft gluons and
consequently its modified fragmentation is more suppressed at intermedium and large $z$
as compared to the quark fragmentation functions. 

At large $z$, the modification factor
for gluon jets is seen to have a small bump for small values of $\hat q$. This
feature results from the competition between gluon bremsstrahlung and pair production in the
evolution of a gluon jet. Both of two processes contribute to the vacuum evolution. In the
medium, however, gluon bremsstrahlung suppresses the fragmentation function while
pair production is relatively enhanced in particular at large $z$. If one switchs off the medium induced $g\rightarrow gg$
and $g \rightarrow q\bar q$ channels, the bump disappears. This feature is similar to the shower parton
distributions of a gluon jet as discussed in the last subsection. However, as one increases the value of $\hat q$,
these features at large $z$ disappear as induced gluon bremsstrahlung further suppresses the gluon
fragmentation functions.

\begin{figure}
  \centering
 \includegraphics[width=\textwidth]{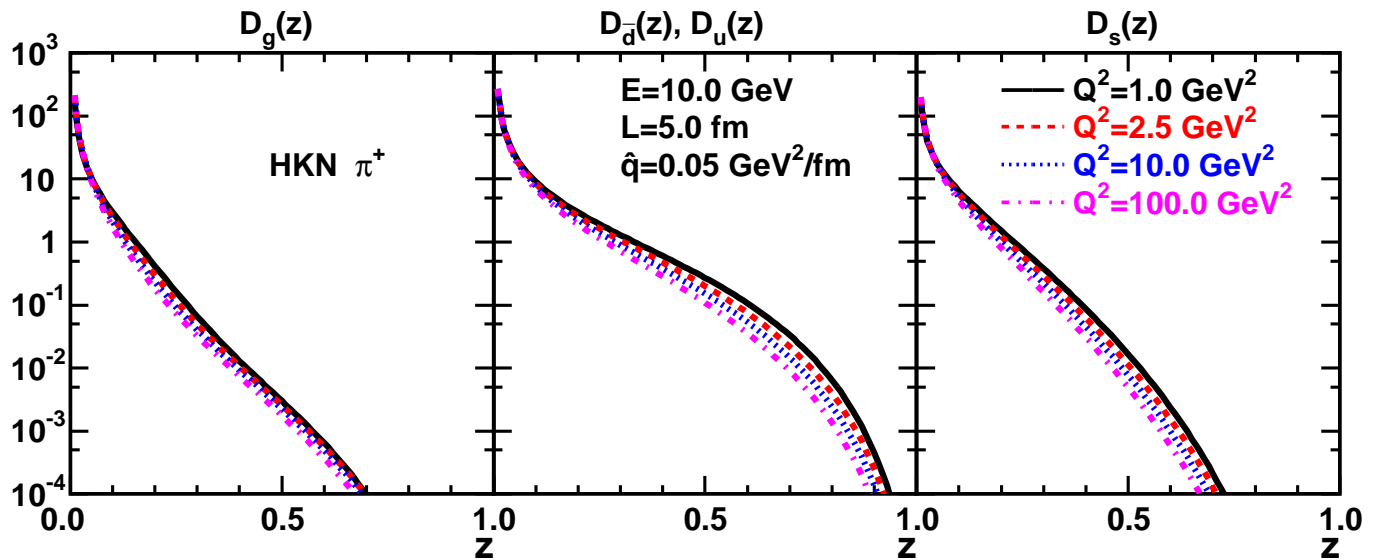}
  \caption{(color online) Modified $\pi$ fragmentation function of a gluon (right), up, down (middle) and strange (left)
  quark jet  with initial energy $E=10$ GeV that propagates through a uniform medium with length $L=5$ fm and jet transport
  parameter $\hat q_{0}=0.05$ GeV$^{2}$/fm. The momentum scale of jet production is $Q^{2}=1.0$ (solid),
  2.5 (dashed) ,10 GeV$^{2}$  (dotted) and 100 GeV (dot-dashed).}
  \label{fig:ffq2}
\end{figure}

\begin{figure}
  \centering
 \includegraphics[width=\textwidth]{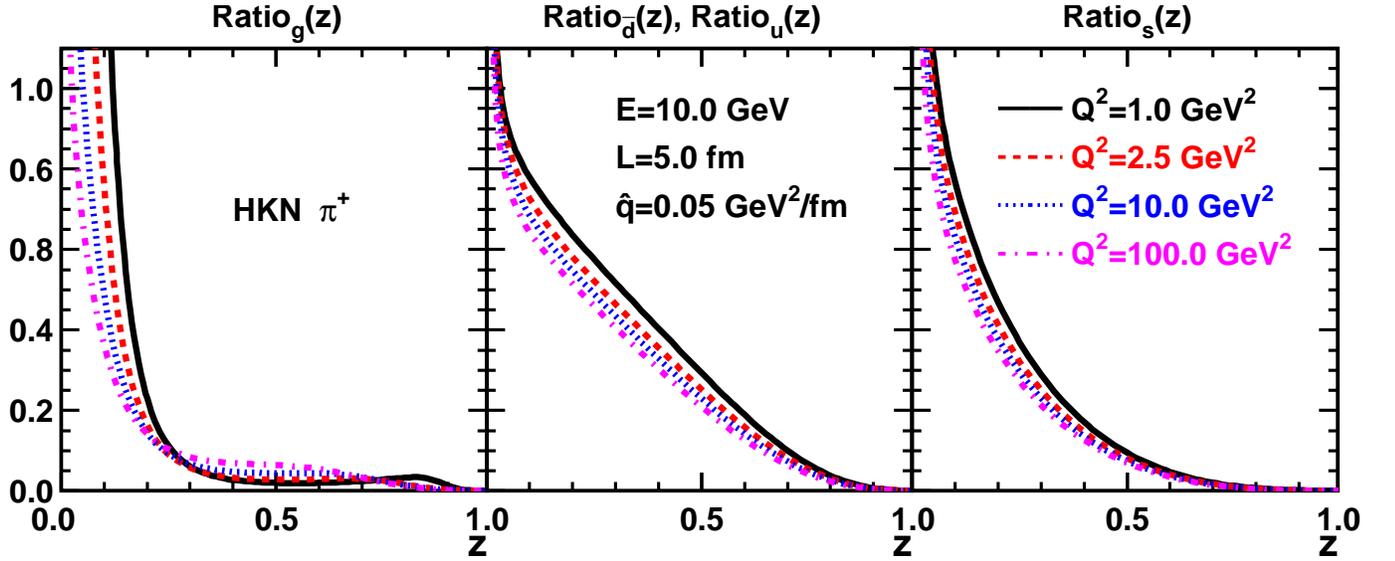}
  \caption{(color online) The modification factor $R_{a}^{\pi}(z,Q^{2})$ for the medium modified fragmentation
  function in Fig.~\ref{fig:ffq2}.}
  \label{fig:rq2}
\end{figure}

\begin{figure}
  \centering
  \includegraphics[width=\textwidth]{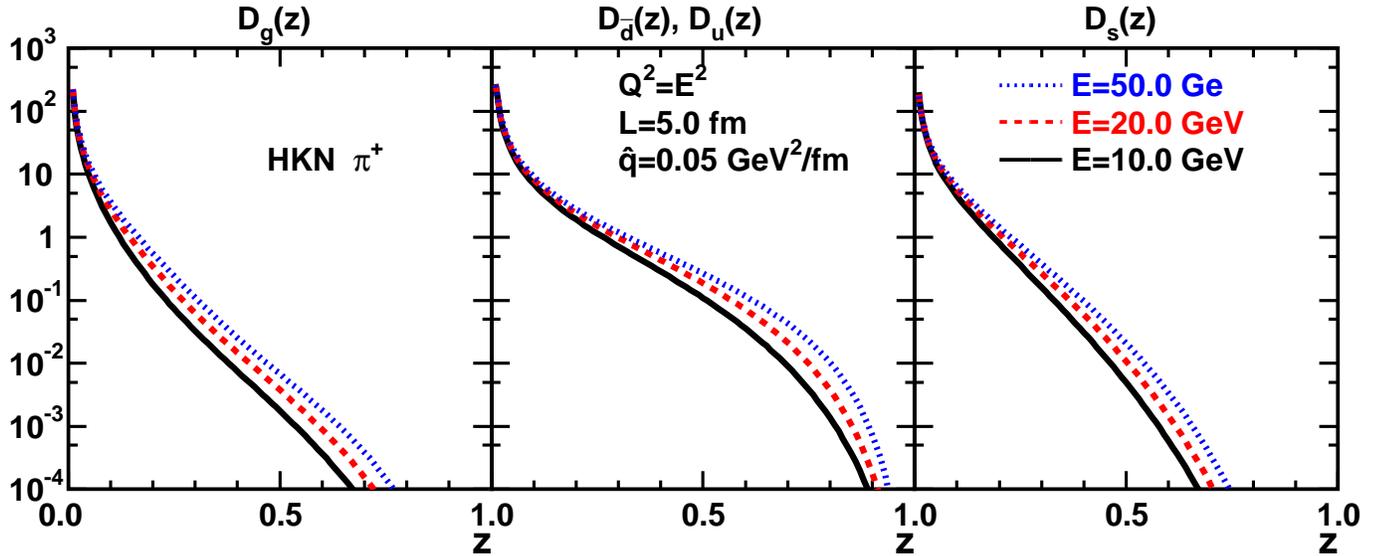}
  \caption{(color online) Modified $\pi$ fragmentation function of a gluon (right), up, down (middle) and strange (left)
  quark jet with initial energy $E=10$ (solid), 20 (dashed) and 50 GeV (dotted) that propagates through
  a uniform medium with length $L=5$ fm and jet transport parameter $\hat q_{0}=0.05$ GeV$^{2}$/fm.
  The momentum scale of jet production is $Q^{2}=E^{2}$.}
  \label{fig:ffe}
\end{figure}

\begin{figure}
  \centering
  \includegraphics[width=\textwidth]{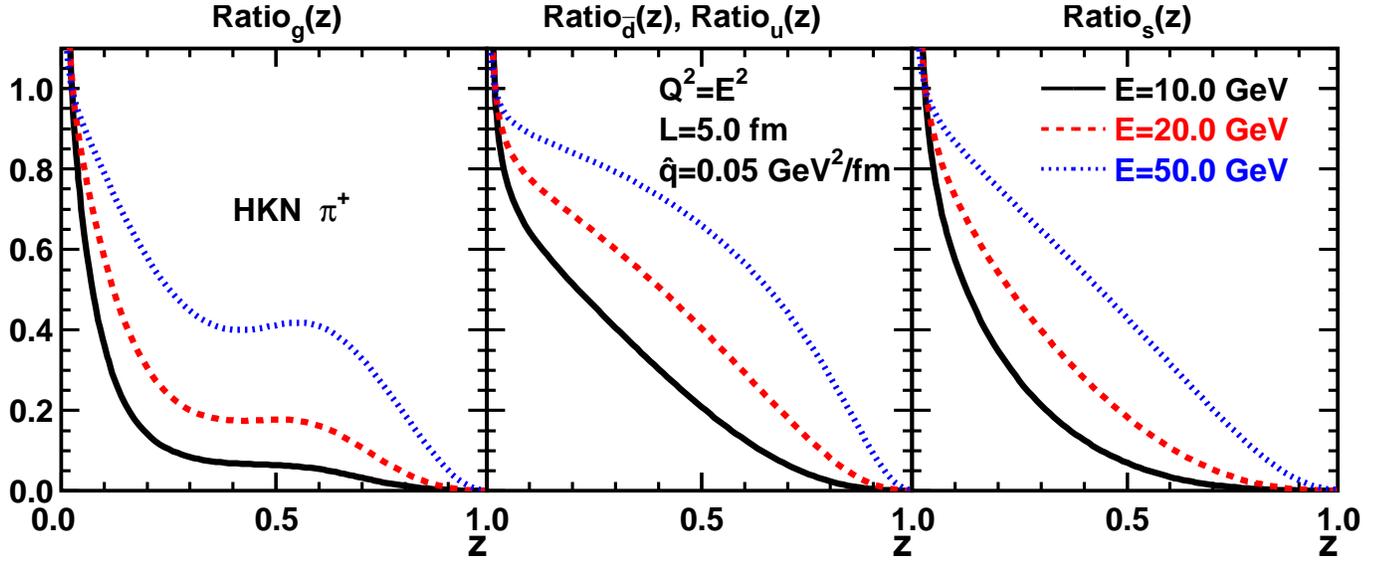}
  \caption{(color online) The modification factor $R_{a}^{\pi}(z,Q^{2})$ for the medium modified fragmentation
  function in Fig.~\ref{fig:ffe}.}
  \label{fig:re}
\end{figure}

Since the modification to the effective splitting functions due to gluon bremsstrahlung induced by
multiple scattering are power suppressed at large $Q^{2}$, most of the the modification to the fragmentation
functions should come from mDGLAP evolution at small $Q^{2}$. Shown in Figs.~\ref{fig:ffq2}
and \ref{fig:rq2} are the modified fragmentation functions and the corresponding modification
factors for different $Q^{2}$ but for fixed initial energy $E$ and value of $\hat q_{0}$=0.05 GeV$^{2}$/fm.
The fragmentation functions at $Q_0^{2}=1$ GeV$^{2}$ correspond to the medium modified initial conditions that we
generate according to our assumed prescription in this paper. One can see in Fig.~\ref{fig:rq2}
that the initial fragmentation functions are already suppressed at large $z$ due to induced gluon emission and therefore
most of the modification comes from mDGLAP evolution near or below the initial value
of $Q^{2}=Q_{0}^{2}$. Medium modification above the initial scale is relatively small. This will give a weak $Q^{2}$
dependence of the medium modification of the fragmentation functions as we will show in the discussion of jet quenching in DIS.
The modification, however, varies with the initial energy $E$ significantly,
as shown in Figs.~\ref{fig:ffe} and \ref{fig:re} for a fixed value of $\hat q_{0}=0.05$ GeV$^{2}$/fm. Such variations reflect the
energy dependence of the energy loss as shown in Fig.~\ref{fig:loss1}. Eventually, when the initial energy
becomes infinitely large, the modification will become increasingly smaller.

 %\begin{figure}
  %\centering
  %\includegraphics[width=0.8\textwidth]{ff_evol_int.eps}
  %\caption{caption}
  %\label{figffe}
%\end{figure}

\section{Modified fragmentation functions in DIS}

\begin{figure}
  \centering
  \includegraphics[width=0.8\textwidth]{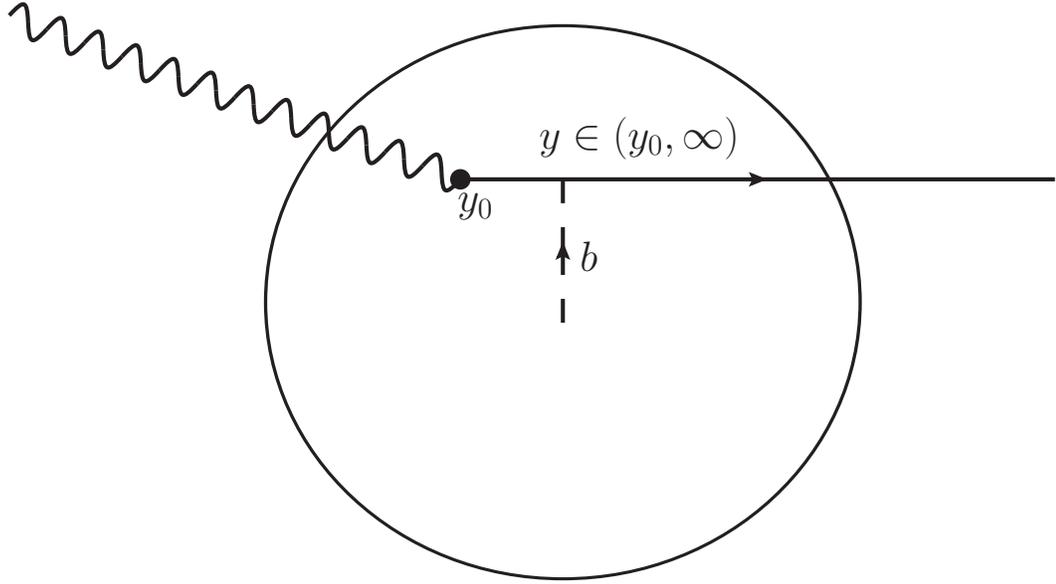}
  \caption{Illustration of the path integration in DIS. See details in text.}
  \label{fig:DIS_illus}
\end{figure}

To calculate the nuclear modified fragmentation functions in semi-inclusive DIS off a nucleus we have to
employ a more realistic form of the nuclear density distribution. We consider the initial quark jet
produced at $y_{0}$ that travels along a direction with impact parameter $b$ (see Fig.~\ref{fig:DIS_illus}
for illustration). We assume that the jet transport parameter along the quark jet trajectory is proportional
 to the nuclear density,
\begin{equation}
\hat q(y,b)=\hat q_{0} \frac{\rho_{A}(y,b)}{\rho_{A}(0,0)},
\end{equation}
where $\rho_{A}(y,b)$ is the nuclear density distribution
normalized as $\int dy d^{2}b\rho_{A}(y,b)=A$ and $\hat q_{0}$ is defined to be the jet transport
parameter at the center of the nucleus. We will
use the three-parameter Wood-Saxon form of nuclear density distribution here.

If we neglect the nuclear and impact parameter dependence of the nuclear quark distribution
function, the photon-nucleon cross section that produces a quark at $(y_{0},b)$ is proportional
to the nuclear density distribution $\rho_{A}(y_{0},b)$. Then the averaged mFF should be
\begin{equation}
 \tilde{D}(z)=\langle\tilde{D}(z,y_0)\rangle=\frac{\pi}{A}\int_{0}^{\infty} db^{2} \int_{-\infty}^{\infty} dy_0 \tilde{D}(z,y_0) \rho_A(y_0,b).
\end{equation}
In order to calculate the $\tilde{D}(y_0,b)$ for a quark produced at location $(y_0,b)$, the path integral in the modified splitting functions should be replaced by the following
\begin{equation}
 \int dy \hat q(y) [\cdots] \rightarrow  \frac{\hat q_{0}}{\rho_{A}(0,0)} \int_{y0}^{\infty} dy\rho_A(y,b)[\cdots]
\end{equation}

%Then the path integral in the modified
%splitting functions should be replaced by the following averaged over the initial production point,
%\begin{equation}
%\int dy \hat q(y) [\cdots] \rightarrow \frac{\hat q_{0}}{A\rho_{A}(0,0)}
%\int d^{2}b \int_{-\infty}^{\infty} dy_{0}\rho_{A}(y_{0},b)\int_{y_{0}}^{\infty} dy \rho_{A}(y,b)[\cdots] .
%\end{equation}

\begin{figure}
  \centering
 \includegraphics[width=0.8\textwidth]{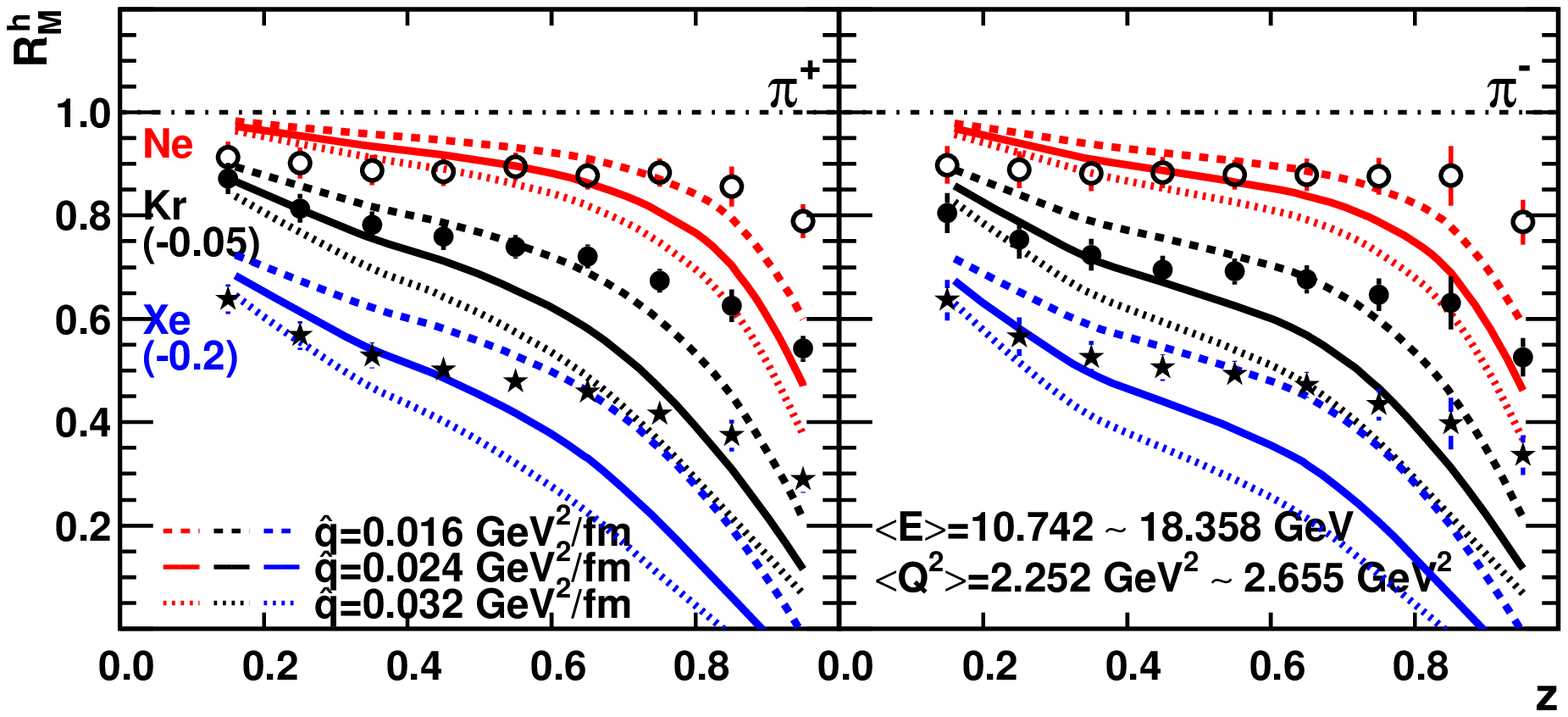}
  \includegraphics[width=0.8\textwidth]{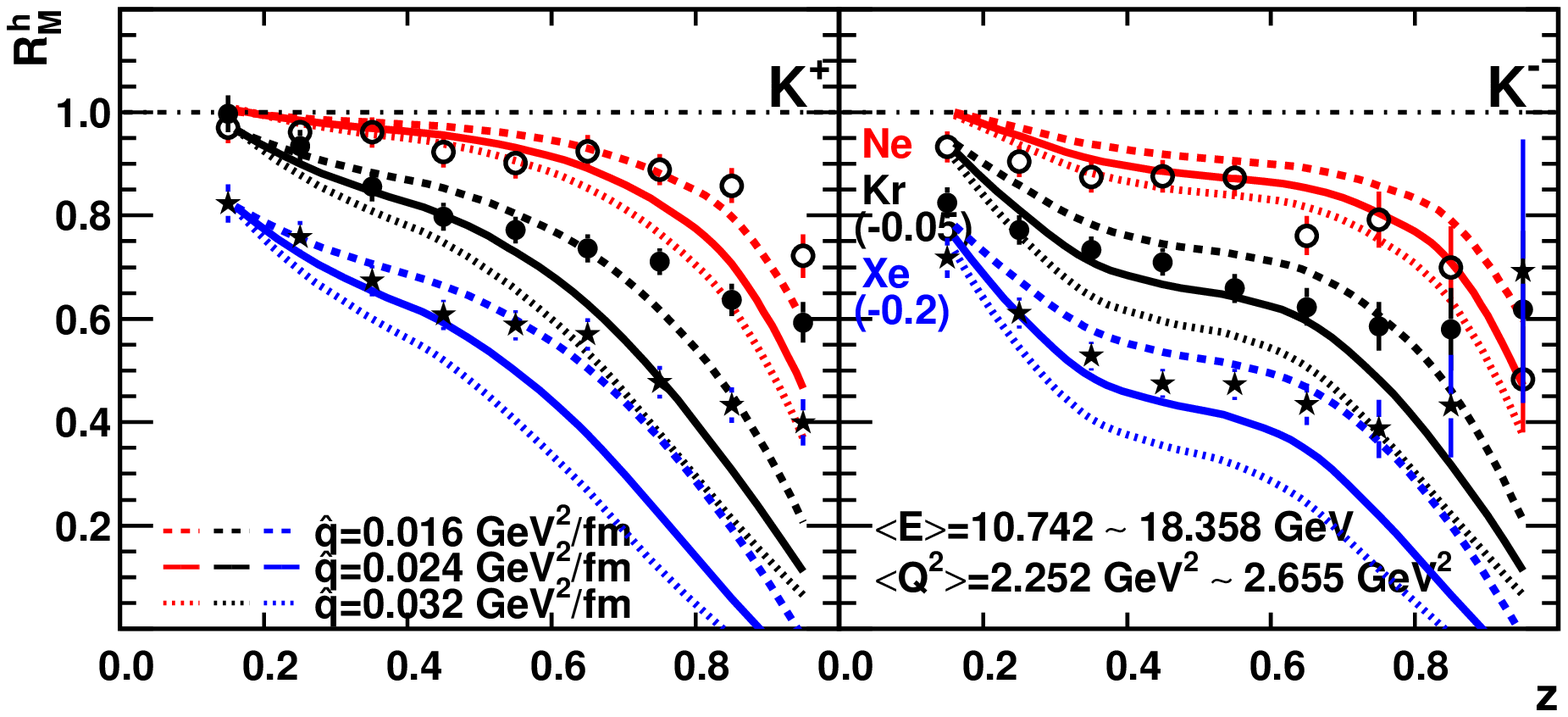}
   \includegraphics[width=0.8\textwidth]{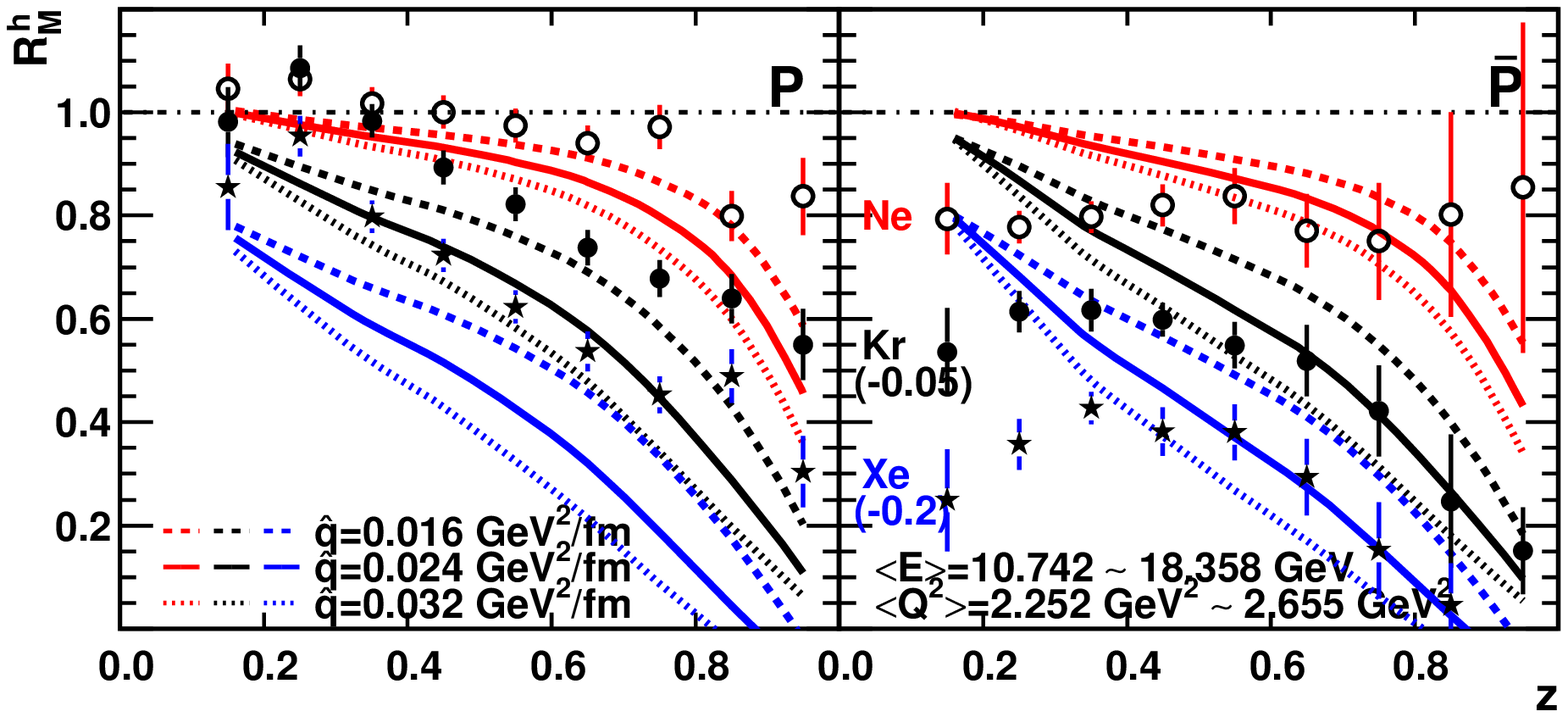}
  \caption{(color online) The modified multiplicity ratios as a function of $z$ with different values of the jet transport parameter $\hat q_{0}$ compared with the HERMES \cite{HERMES} data for $Ne$,$Kr$ and $Xe$ targets. For clear presentation the modification factors for different targets have been shifted vertically by
  some value($Kr$ by -0.05 and $Xe$ by -0.2).}
  \label{fig:HERMES_z}
\end{figure}

Shown in Fig.~\ref{fig:HERMES_z} are the calculated nuclear modification factor of $\pi^{\pm}$, $K^{\pm}$
and $p(\bar p)$ fragmentation functions with different values of jet transport parameter $\hat q_{0}=0.016 - 0.032$
GeV$^{2}$ as compared to the HERMES experimental
data \cite{HERMES}. The experimental results are
presented in terms of multiplicity ratio $R_M^h$, which represents the ratio of the number
of hadrons of type $h$ produced per DIS event for a nuclear target
of mass A to that for a deuterium target \cite{hadron-ratio}:
\begin{eqnarray}
\nonumber
R_M^h(z,\nu)&=&\left(\frac{N^h(z,\nu)}{N^e(\nu)}|_A\right) / \left(\frac{N^h(z,\nu)}{N^e(\nu)}|_D\right)\\
&=&\left(\frac{\Sigma e_f^2q_f(x)D_f^h(z)}{\Sigma e_f^2q_f(x)}|_A\right) / \left(\frac{\Sigma e_f^2q_f(x)D_f^h(z)}{\Sigma e_f^2q_f(x)}|_D\right).
\end{eqnarray}
In the experimental data, each $z$ bin has an averaged value of initial jet energy $\langle E\rangle=\langle \nu\rangle$
and virtuality $\langle Q^{2}\rangle$ which we use in the calculation. Since the averaged values of $Q^{2}=2.25-2.65$ GeV$^{2}$ in HERMES data are quite small, the suppression factors are apparently sensitive to the value of the initial
evolution scale $Q^2_0$.  In the calculation, we have used
$Q^2_0=1.0$ GeV$^{2}$. We used the HKN \cite{HKN} parameterizations as initial condition which extends to
down to scale $Q^{2}=1$ GeV$^{2}$.
The agreement between the calculation from
mDGLAP evolution and the HERMES experimental data \cite{HERMES} are quite well for three nuclear targets
in the intermediate region of $z$ values.  At large values of $z$ the agreement is not so well maybe due to other
effects such as hadronic interaction \cite{hadronic,Arleo:2003jz} that we have not taken into account.  The calculated
modification factors for protons in a large nucleus are also much smaller than the experimental data. This
might be related to the non-perturbative baryon transport in hadronic processes \cite{Kharzeev:1996sq}.
We have so far neglected quark-anti-quark annihilation contribution to the mDGLAP evolution equations.
These processes will affect the medium modification of anti-quarks and will likely improve the modification
factor for anti-proton distribution.

\begin{figure}
  \centering
 \includegraphics[width=0.8\textwidth]{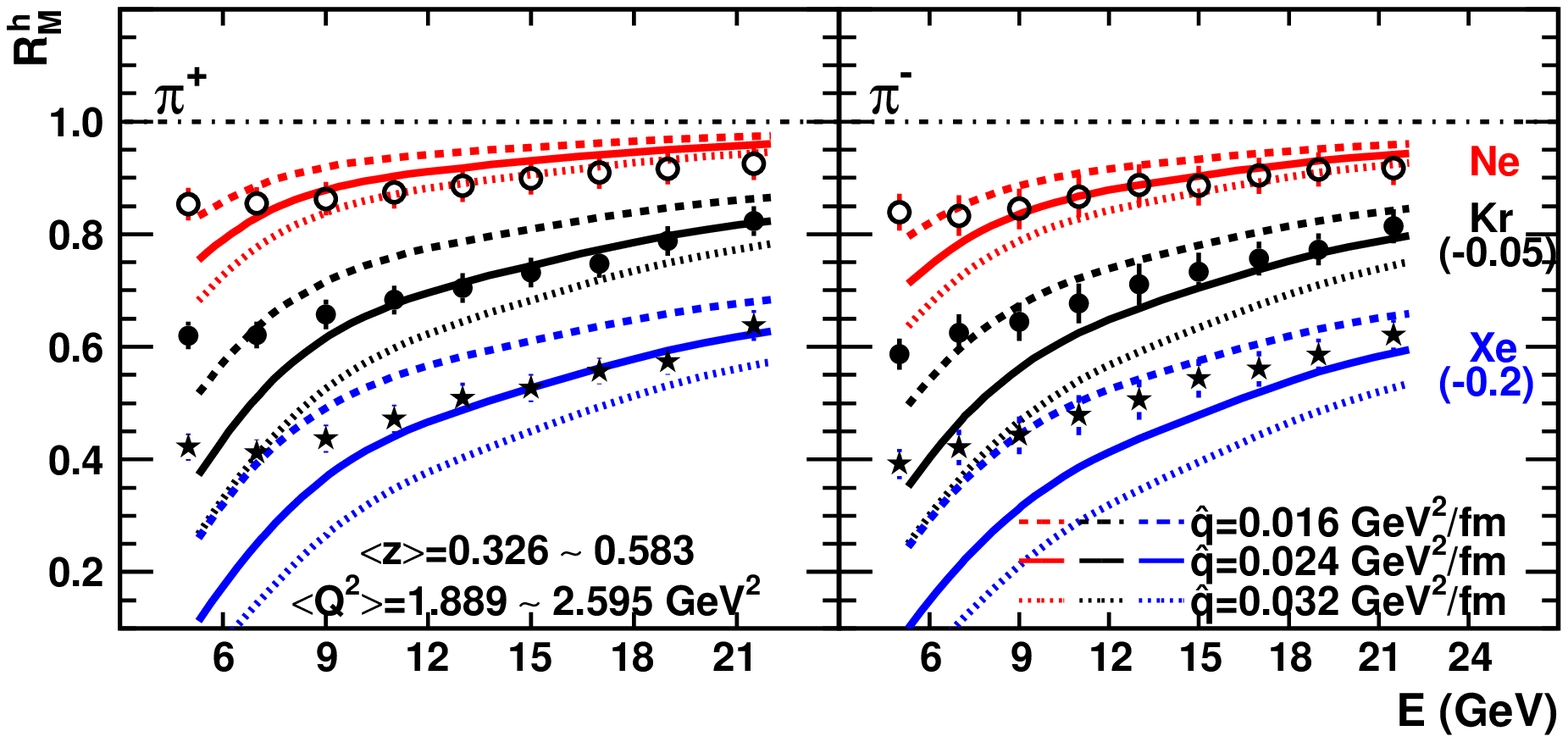}
 \includegraphics[width=0.8\textwidth]{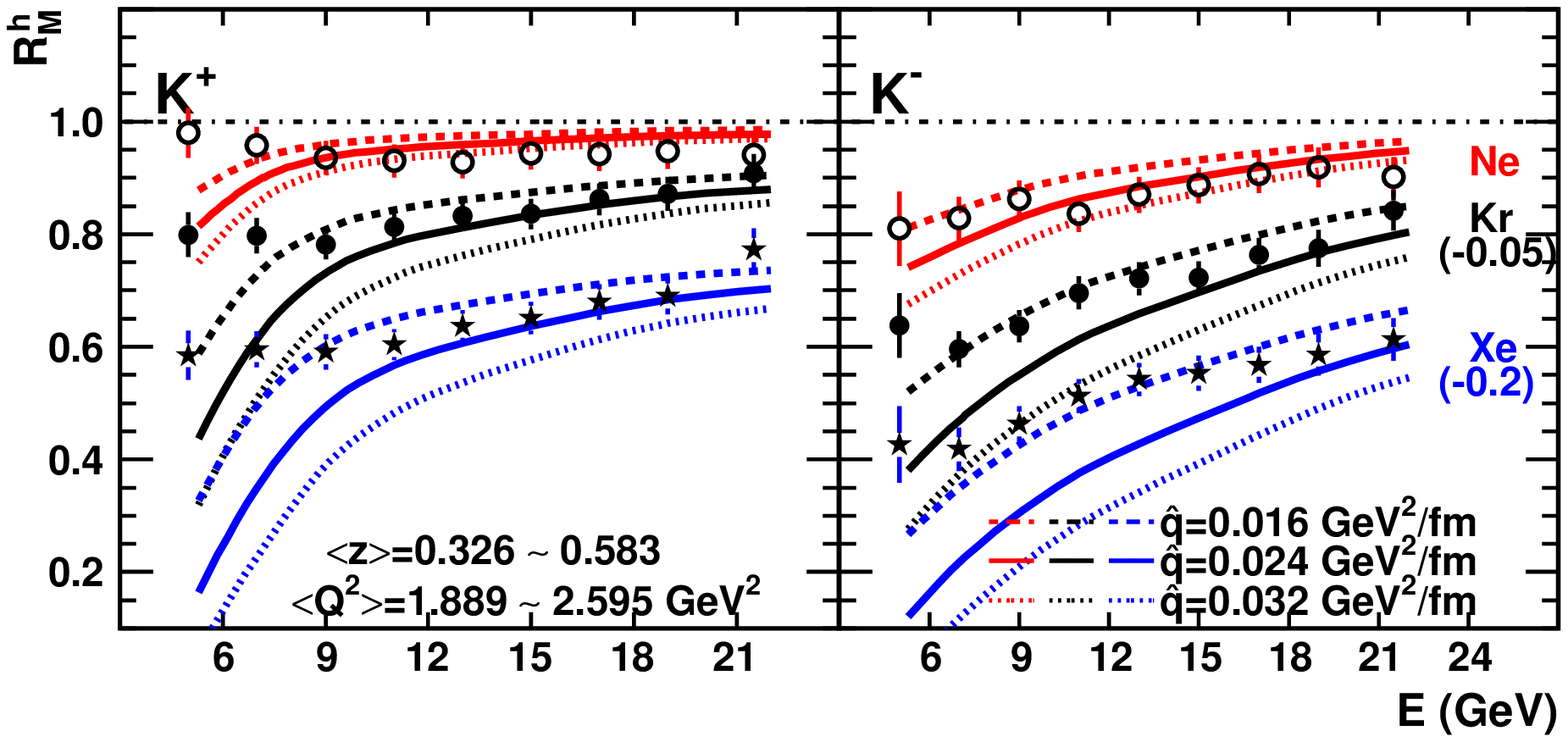}
 \includegraphics[width=0.8\textwidth]{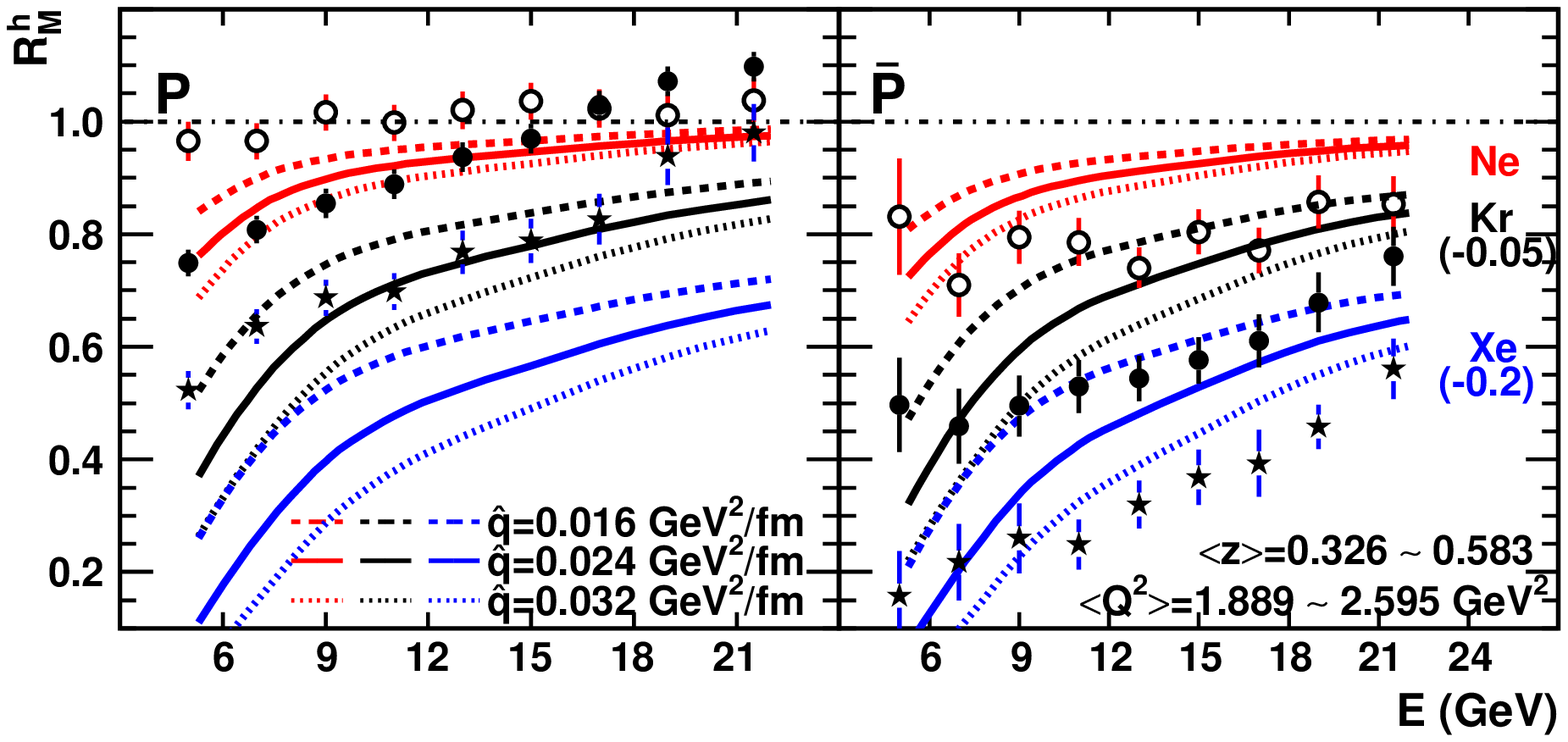}
  \caption{(color online) The energy dependency of the nuclear modification factors with different values of the jet transport
  parameter $\hat q_{0}$ compared with the HERMES \cite{HERMES}
  data for $Ne$, $Kr$ and $Xe$ targets. For clear presentation the modification factors for different targets have
  been shifted vertically by some value($Kr$ by -0.05 and $Xe$ by -0.2).}
  \label{fig:HERMES_nu}
\end{figure}

Shown in Fig.~(\ref{fig:HERMES_nu}) are the calculated nuclear modification factors at fixed $z$ as
a function of the initial jet energy $E$ as compared to the HERMES experimental data \cite{HERMES}.
Again, each bin of $E$ has an average value of $\langle z \rangle$ and $\langle Q^{2}\rangle$ which
we also use in the calculation. Similarly as illustrated in the uniform medium (a ``brick''), the medium
modification of the fragmentation function gradually disappears as the initial jet energy $E$ increases.
The agreement between our theory calculations and experimental data is generally good for a range of
values of $\hat q_{0}=0.016-0.032$ GeV$^{2}$, except at lower energy where hadronic absorption
might become important. From the combined fit we find the jet transport parameter at the center
of large nuclei is $\hat q_{0}\approx 0.024\pm 0.008$ GeV$^{2}$/fm. This value is consistent with the transverse
momentum broadening of the Drell-Yan dilepton production in $p+A$ collisions \cite{Guo:1998rd,Guo:1999eh,dy}
 $\langle \Delta q_{T}^{2}\rangle=0.016 A^{1/3}$ GeV$^{2}$ which gives an averaged jet transport
 parameter $\hat q_{0}=0.018$ GeV$^{2}$/fm.

\begin{figure}
  \centering
 \includegraphics[width=0.8\textwidth]{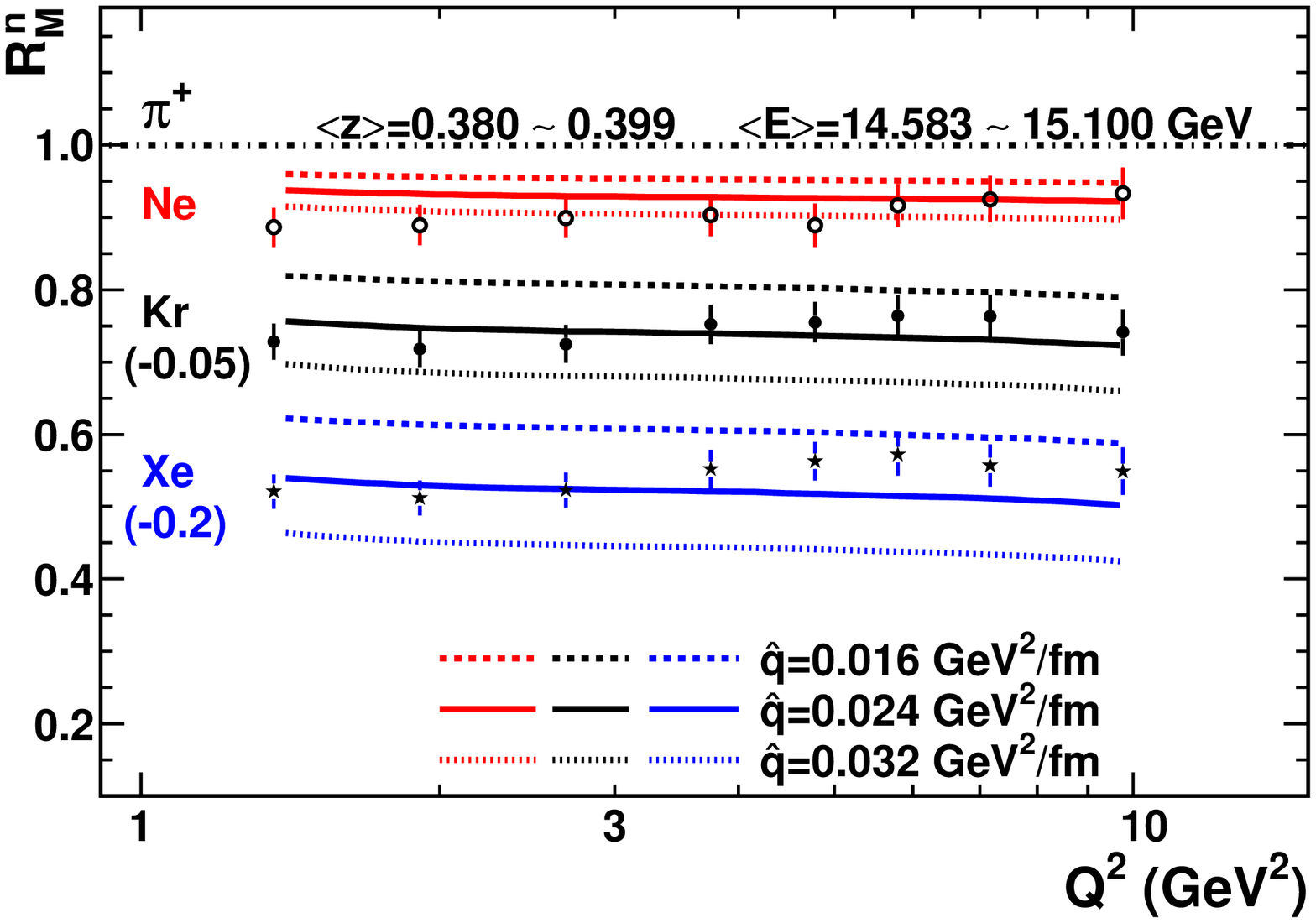}
  \caption{(color online) Comparison of the modified multiplicity ratios as a function of $Q^{2}$ at fixed value of $z$ and jet energy $E$ with the HERMES \cite{HERMES} data for $Ne$, $Kr$ and $Xe$ targets. For clear presentation the modification factors for
  different targets have been shifted vertically by some value($Kr$ by -0.05 and $Xe$ by -0.2).}
  \label{fig:HERMES_q}
\end{figure}

As described in Sec. \ref{sub-initial}, we have modeled the initial condition for fragmentation functions
in the medium at $Q_{0}^{2}=1$ GeV$^{2}$ as different from the fragmentation functions in vacuum
due to energy loss of parton nearly on shell.
Therefore, most of the medium modification to the fragmentation functions come from mDGLAP
evolution at low $Q^{2}$ while contribution from high $Q^{2}$ region is power-suppressed. This will lead
to a very weak $Q^{2}$ dependence as shown in Fig.~\ref{fig:HERMES_q}, where we compared
the calculated modification factors for the fragmentation function with experimental data
at fixed $z$ and initial jet energy $E$ but as a function of $Q^{2}$. The calculated suppression factors are
almost independent of $Q^{2}$, consistent with the experimental data \cite{HERMES}. If one has chosen
the initial condition for medium modified fragmentation functions at $Q_{0}^{2}$ as the same as the
vacuum one, one would obtain a modification factor that has a too strong $Q^{2}$ dependence to be
consistent with the experimental data.

\section{Summary}

In this paper, we have extended the modified DGLAP evolution equations to include induced gluon radiation for gluon jet
and quark-anti-quark pair creation from gluon fusion \cite{Schafer:2007xh} within the framework of generalized
factorization for higher-twist contribution to multiple parton scattering. The effective parton splitting functions
are proportional to a path integration of the jet transport parameter $\hat q$ over the propagation length.
We then numerically solve the coupled mDGLAP
equations for medium modified fragmentation functions for different static profile of medium and different
values of the jet transport parameter $\hat q$. For a ``brick'' profile,
we have studied the shower parton distributions within a jet during its propagation through the medium and relate
the change of momentum fraction carried by the valence quark as the effective energy loss of a propagating quark.
We have systematically studied the energy, medium length and virtuality dependence of the medium modified
fragmentation functions. Because of the inclusion of induced pair creation and different splitting functions for gluon
and quark jets in the mDGLAP evolution, we found that the modification factors are different for gluon, valence
quark and sea quark fragmentation functions.

We also applied the mDGLAP evolution to quark propagation in the deeply
inelastic scattering (DIS) of a large nucleus and found the calculated nuclear modification of the effective
fragmentation functions in good agreement with experimental data in the intermediate $z$ region.  In modeling the initial condition for
modified fragmentation functions, we have chosen to include medium induced radiation and parton
energy loss below the initial scale $Q_{0}^{2}$. In this case, most of the medium modification
comes from mDGLAP evolution in the low $Q^{2}$ region while large $Q^{2}$ contribution is
power-suppressed. This leads to a weak $Q^{2}$ dependence of the medium modification of the
fragmentation functions which is consistent with the experimental data in DIS.

\section*{Acknowledgement}

We would like to acknowledge helpful discussions with A. Majumder. X.N.W thanks the hospitality
of the Physics Department of Shandong University during the completion of this work.
This work is supported  by the Director, Office of Energy
Research, Office of High Energy and Nuclear Physics, Divisions of
Nuclear Physics, of the U.S. Department of Energy under Contract No.
DE-AC02-05CH11231 and National Natural Science Foundation of China
under Project Nos. 10525523.

%\end{multicols}


\begin{thebibliography}{88}

\bibitem{wg90}
  X.~N.~Wang and M.~Gyulassy,
  %``Gluon shadowing and jet quenching in A + A collisions at s**(1/2) =
  %200-GeV,''
  Phys.\ Rev.\ Lett.\  {\bf 68}, 1480 (1992).
  %%CITATION = PRLTA,68,1480;%%

\bibitem{Gyulassy:1993hr}
  M.~Gyulassy and X.~N.~Wang,
  %``Multiple collisions and induced gluon Bremsstrahlung in QCD,''
  Nucl.\ Phys.\  B {\bf 420}, 583 (1994).
  %[arXiv:nucl-th/9306003].
  %%CITATION = NUPHA,B420,583;%%

\bibitem{Baier:1996sk}
  R.~Baier, Y.~L.~Dokshitzer, A.~H.~Mueller, S.~Peigne and D.~Schiff,
  %``Radiative energy loss and p(T)-broadening of high energy partons in
  %nuclei,''
  Nucl.\ Phys.\  B {\bf 484}, 265 (1997)
  %[arXiv:hep-ph/9608322].
  %%CITATION = NUPHA,B484,265;%%

\bibitem{Wiedemann:2000za}
  U.~A.~Wiedemann,
  %``Gluon radiation off hard quarks in a nuclear environment: Opacity
  %expansion,''
  Nucl.\ Phys.\  B {\bf 588}, 303 (2000)
  %[arXiv:hep-ph/0005129].
  %%CITATION = NUPHA,B588,303;%%

\bibitem{Gyulassy:2000er}
M.~Gyulassy, P.~Levai and I.~Vitev,
  %``Reaction operator approach to non-Abelian energy loss,''
  Nucl.\ Phys.\  B {\bf 594}, 371 (2001)
 % [arXiv:nucl-th/0006010].
  %%CITATION = NUPHA,B594,371;%%

  \bibitem{Guo:2000nz}
  X.~F.~Guo and X.~N.~Wang,
  %``Multiple Scattering, Parton Energy Loss and Modified Fragmentation
  %Functions in Deeply Inelastic eA Scattering,''
  Phys.\ Rev.\ Lett.\  {\bf 85}, 3591 (2000)
  %[arXiv:hep-ph/0005044].
  %%CITATION = PRLTA,85,3591;%%


\bibitem{Wang:2001ifa}
  X.~N.~Wang and X.~F.~Guo,
  %``Multiple parton scattering in nuclei: Parton energy loss,''
  Nucl.\ Phys.\  A {\bf 696}, 788 (2001)
  %[arXiv:hep-ph/0102230].
  %%CITATION = NUPHA,A696,788;%%

  \bibitem{Wang:1998bha}
  X.~N.~Wang,
  %``Effect of jet quenching on high $p_{T}$ hadron spectra in high-energy
  %nuclear collisions,''
  Phys.\ Rev.\  C {\bf 58}, 2321 (1998)
  %[arXiv:hep-ph/9804357].
  %%CITATION = PHRVA,C58,2321;%%

\bibitem{Wang:1998ww}
  X.~N.~Wang,
  %``Systematic study of high $p_{T}$ hadron spectra in $p p$, $p$ A and A A
  %collisions from SPS to RHIC energies,''
  Phys.\ Rev.\  C {\bf 61}, 064910 (2000)
  %[arXiv:nucl-th/9812021].
  %%CITATION = PHRVA,C61,064910;%%

\bibitem{Wang:1996yh}
  X.~N.~Wang, Z.~Huang and I.~Sarcevic,
  %``Jet quenching in the opposite direction of a tagged photon in  high-energy
  %heavy-ion collisions,''
  Phys.\ Rev.\ Lett.\  {\bf 77}, 231 (1996)
  %[arXiv:hep-ph/9605213].
  %%CITATION = PRLTA,77,231;%%

  \bibitem{Wang:1996pe}
  X.~N.~Wang and Z.~Huang,
  %``Medium-induced parton energy loss in gamma + jet events of  high-energy
  %heavy-ion collisions,''
  Phys.\ Rev.\  C {\bf 55}, 3047 (1997)
  %[arXiv:hep-ph/9701227].
  %%CITATION = PHRVA,C55,3047;%%

\bibitem{arleo}
  F.~Arleo,
  %``Hard pion and prompt photon at RHIC, from single to double inclusive
  %production,''
  JHEP {\bf 0609}, 015 (2006).
  %[arXiv:hep-ph/0601075].
  %%CITATION = JHEPA,0609,015;%%



\bibitem{Renk:2006qg}
  T.~Renk,
  %``Towards jet tomography: gamma hadron correlations,''
  Phys.\ Rev.\  C {\bf 74}, 034906 (2006)
  %[arXiv:hep-ph/0607166].
  %%CITATION = PHRVA,C74,034906;%%


\bibitem{zoww09}
  H.~Zhang, J.~F.~Owens, E.~Wang and X.~N.~Wang,
  %``Tomography of high-energy nuclear collisions with photon-hadron
  %correlations,''
  Phys.\ Rev.\ Lett.\  {\bf 103}, 032302 (2009)
  %[arXiv:0902.4000 [nucl-th]].
  %%CITATION = PRLTA,103,032302;%%


  \bibitem{Qin:2009bk}
  G.~Y.~Qin, J.~Ruppert, C.~Gale, S.~Jeon and G.~D.~Moore,
  %``Jet energy loss, photon production, and photon-hadron correlations at
  %RHIC,''
  Phys.\ Rev.\  C {\bf 80}, 054909 (2009)
  %[arXiv:0906.3280 [hep-ph]].
  %%CITATION = PHRVA,C80,054909;%%



\bibitem{Wang:2003mm}
  X.~N.~Wang,
  %``High-$p_T$ Hadron Spectra, Azimuthal Anisotropy and Back-to-Back
  %Correlations in High-energy Heavy-ion Collisions,''
  Phys.\ Lett.\  B {\bf 595}, 165 (2004)
  %[arXiv:nucl-th/0305010].
  %%CITATION = PHLTA,B595,165;%%

\bibitem{zoww07}
  H.~Z.~Zhang, J.~F.~Owens, E.~Wang and X.~N.~Wang,
  %``Dihadron Tomography of High-Energy Nuclear Collisions in NLO pQCD''
  Phys.\ Rev.\ Lett.\  {\bf 98}, 212301 (2007).
  %arXiv:nucl-th/0701045
  %%CITATION = PRLTA,98,212301;%%

  \bibitem{Majumder:2004pt}
  A.~Majumder, E.~Wang and X.~N.~Wang,
  %``Modified dihadron fragmentation functions in hot and nuclear matter,''
  Phys.\ Rev.\ Lett.\  {\bf 99}, 152301 (2007)
  %[arXiv:nucl-th/0412061].
  %%CITATION = PRLTA,99,152301;%%

\bibitem{phenix} K.~Adcox {\it et al.},
  %``Suppression of hadrons with large transverse momentum in central  Au + Au
  %collisions at s**(1/2)(N N) = 130-GeV,''
  Phys.\ Rev.\ Lett.\  {\bf 88}, 022301 (2002).
  %%CITATION = NUCL-EX 0109003;%%

\bibitem{star0}
  C.~Adler {\it et al.}  ,
  %[STAR Collaboration],
  %``Centrality dependence of high p(T) hadron suppression in Au + Au collisions
  %at s(NN)**(1/2) = 130-GeV,''
  Phys.\ Rev.\ Lett.\  {\bf 89}, 202301 (2002).
%  [arXiv:nucl-ex/0206011].
  %%CITATION = PRLTA,89,202301;%%

\bibitem{star} C.~Adler {\it et al.},
  %``Disappearance of back-to-back high p(T) hadron correlations in central Au +
  %Au collisions at s(NN)**(1/2) = 200-GeV,''
  Phys.\ Rev.\ Lett.\  {\bf 90}, 082302 (2003).
  %%CITATION = NUCL-EX 0210033;%%

\bibitem{star-gam-hadr}
J.~Adams {\it et al.} ,
%[STAR Collaboration],
  %``Direct observation of dijets in central Au + Au collisions at  s(NN)**(1/2)
  %= 200-GeV,''
  Phys.\ Rev.\ Lett.\  {\bf 97}, 162301 (2006);
%  [arXiv:nucl-ex/0604018].
  %%CITATION = PRLTA,97,162301;%%
A.~M.~Hamed,
  %``Direct photon-charged hadron azimuthal correlations,''
  J.\ Phys.\ G {\bf 35}, 104120 (2008);
  %[arXiv:0806.2190 [nucl-ex]].
  %%CITATION = JPHGB,G35,104120;%%
  %``Modified Fragmentation Function in Heavy Ion Collisions at RHIC via Direct photon-Jet Measurements''
  %arXiv:0809.1462 [nucl-ex].
  %%CITATION = ARXIV:0809.1462;%%

\bibitem{frantz}
  J.~Frantz,
  %``Two-particle Direct Photon-Jet Correlation Measurements in PHENIX,''
  arXiv:0901.1393 [nucl-ex].
  %%CITATION = ARXIV:0901.1393;%%


  \bibitem{Vitev:2002pf}
  I.~Vitev and M.~Gyulassy,
  %``High-p(T) tomography of d + Au and Au + Au at SPS, RHIC, and LHC,''
  Phys.\ Rev.\ Lett.\  {\bf 89}, 252301 (2002)
  %[arXiv:hep-ph/0209161].
  %%CITATION = PRLTA,89,252301;%%

  \bibitem{Eskola:2004cr}
  K.~J.~Eskola, H.~Honkanen, C.~A.~Salgado and U.~A.~Wiedemann,
  %``The fragility of high-p(T) hadron spectra as a hard probe,''
  Nucl.\ Phys.\  A {\bf 747}, 511 (2005)
  %[arXiv:hep-ph/0406319].
  %%CITATION = NUPHA,A747,511;%%

\bibitem{Turbide:2005fk}
  S.~Turbide, C.~Gale, S.~Jeon and G.~D.~Moore,
  %``Energy loss of leading hadrons and direct photon production in evolving
  %quark-gluon plasma,''
  Phys.\ Rev.\  C {\bf 72}, 014906 (2005)
  %[arXiv:hep-ph/0502248].
  %%CITATION = PHRVA,C72,014906;%%

\bibitem{majumder}
  For a comparison of different models and their phenomenology, see
  A. Majumder,
  %``A comparative study of jet-quenching schemes,''
  arXiv:nucl-th/0702066;
  %%CITATION = NUCL-TH/0702066;%%
  to be published in the proceedings of Quark Matter 2006,
  Shanghai, Nov. 14-19, 2006.



\bibitem{ww02} E.~Wang and X.-N.~Wang,
%``Jet tomography of dense and nuclear matter,''
  Phys.\ Rev.\ Lett.\
  {\bf 89}, 162301 (2002).
  %%CITATION = HEP-PH 0202105;%%

  \bibitem{Aurenche:1999nz}
  P.~Aurenche, M.~Fontannaz, J.~P.~Guillet, B.~A.~Kniehl and M.~Werlen,
  %``Large p(T) inclusive pi0 cross-sections and next-to-leading-order QCD
  %predictions,''
  Eur.\ Phys.\ J.\  C {\bf 13}, 347 (2000)
  %[arXiv:hep-ph/9910252].
  %%CITATION = EPHJA,C13,347;%%

  \bibitem{Owens:2001rr}
  J.~F.~Owens,
  %``A next-to-leading-order study of dihadron production,''
  Phys.\ Rev.\  D {\bf 65}, 034011 (2002)
  %[arXiv:hep-ph/0110036].
  %%CITATION = PHRVA,D65,034011;%%

  \bibitem{Baer:1990ra}
  H.~Baer, J.~Ohnemus and J.~F.~Owens,
  %``A NEXT-TO-LEADING LOGARITHM CALCULATION OF DIRECT PHOTON PRODUCTION,''
  Phys.\ Rev.\  D {\bf 42}, 61 (1990).
  %%CITATION = PHRVA,D42,61;%%

 \bibitem{dglap}
  % \bibitem{Dokshitzer:1977sg}
  Y.~L.~Dokshitzer,
  %``Calculation Of The Structure Functions For Deep Inelastic Scattering And E+
  %E- Annihilation By Perturbation Theory In Quantum Chromodynamics,''
  Sov.\ Phys.\ JETP {\bf 46}, 641 (1977)
  [Zh.\ Eksp.\ Teor.\ Fiz.\  {\bf 73}, 1216 (1977)].
  %%CITATION = ZETFA,73,1216;%%
  V.~N.~Gribov and L.~N.~Lipatov,
  %``Deep Inelastic E P Scattering In Perturbation Theory,''
  Sov.\ J.\ Nucl.\ Phys.\  {\bf 15}, 438 (1972)
  [Yad.\ Fiz.\  {\bf 15}, 781 (1972)].
  %%CITATION = YAFIA,15,781;%%
   %\bibitem{Altarelli:1977zs}
  G.~Altarelli and G.~Parisi,
  %``Asymptotic Freedom In Parton Language,''
  Nucl.\ Phys.\  B {\bf 126}, 298 (1977).
  %%CITATION = NUPHA,B126,298;%%

 \bibitem{fields}
   R. D. Field, \textit{Application of Perturbative QCD}, Vol. 77 of Frontiers in Physics Lecture Series (Addison-Wesley, Reading, MA, 1989).

   \bibitem{Armesto:2007dt}
  N.~Armesto, L.~Cunqueiro, C.~A.~Salgado and W.~C.~Xiang,
  %``Medium-evolved fragmentation functions,''
  JHEP {\bf 0802}, 048 (2008)
  %[arXiv:0710.3073 [hep-ph]].
  %%CITATION = JHEPA,0802,048;%%


\bibitem{Zapp:2008af}
  K.~Zapp, J.~Stachel and U.~A.~Wiedemann,
  %``A local Monte Carlo implementation of the non-abelian
  %Landau-Pomerantschuk-Migdal effect,''
  arXiv:0812.3888 [hep-ph].
  %%CITATION = ARXIV:0812.3888;%%
\bibitem{Majumder:2009zu}
  A.~Majumder,
  %``The in-medium scale evolution in jet modification,''
  arXiv:0901.4516 [nucl-th].
  %%CITATION = ARXIV:0901.4516;%%


\bibitem{Zhang:2003yn}
  B.~W.~Zhang and X.~N.~Wang,
  %``Multiple parton scattering in nuclei: Beyond helicity amplitude
  %approximation,''
  Nucl.\ Phys.\  A {\bf 720}, 429 (2003)
  %[arXiv:hep-ph/0301195].
  %%CITATION = NUPHA,A720,429;%%

\bibitem{Landau:1953gr}
  L.~D.~Landau and I.~Pomeranchuk,
  %``Electron cascade process at very high-energies,''
  Dokl.\ Akad.\ Nauk Ser.\ Fiz.\  {\bf 92} (1953) 735.
  %%CITATION = DANKA,92,735;%%


\bibitem{Migdal:1956tc}
  A.~B.~Migdal,
  %``Bremsstrahlung And Pair Production In Condensed Media At High-Energies,''
  Phys.\ Rev.\  {\bf 103}, 1811 (1956).
  %%CITATION = PHRVA,103,1811;%%


%\cite{Osborne:2002st}
\bibitem{Osborne:2002st}
  J.~Osborne and X.~N.~Wang,
  %``Multiple parton scattering in nuclei: Twist-four nuclear matrix  elements
  %and off-forward parton distributions,''
  Nucl.\ Phys.\  A {\bf 710}, 281 (2002)
  %[arXiv:hep-ph/0204046].
  %%CITATION = NUPHA,A710,281;%%

%\cite{CasalderreySolana:2007sw}
\bibitem{CasalderreySolana:2007sw}
  J.~Casalderrey-Solana and X.~N.~Wang,
  %``Energy dependence of jet transport parameter and parton saturation in
  %quark-gluon plasma,''
  Phys.\ Rev.\  C {\bf 77}, 024902 (2008)
  %[arXiv:0705.1352 [hep-ph]].
  %%CITATION = PHRVA,C77,024902;%%


%\cite{Wang:2006qr}
\bibitem{Wang:2006qr}
  X.~N.~Wang,
  %``Interference effect in elastic parton energy loss in a finite medium,''
  Phys.\ Lett.\  B {\bf 650}, 213 (2007)
  %[arXiv:nucl-th/0604040].
  %%CITATION = PHLTA,B650,213;%%


\bibitem{Schafer:2007xh}
  A.~Schafer, X.~N.~Wang and B.~W.~Zhang,
  %``Multiple Parton Scattering in Nuclei: Quark-quark Scattering,''
  Nucl.\ Phys.\  A {\bf 793}, 128 (2007)
  %[arXiv:0704.0106 [hep-ph]].
  %%CITATION = NUPHA,A793,128;%%


 \bibitem{hoppet}
 %\bibitem{Salam:2008qg}
  G.~P.~Salam and J.~Rojo,
  %``A Higher Order Perturbative Parton Evolution Toolkit (HOPPET),''
  Comput.\ Phys.\ Commun.\  {\bf 180}, 120 (2009)
  %[arXiv:0804.3755 [hep-ph]].
  %%CITATION = CPHCB,180,120;%%


\bibitem{LQS}
 %\bibitem{Luo:1992fz}
  M.~Luo, J.~W.~Qiu and G.~Sterman,
  %``Nuclear dependence at large transverse momentum,''
  Phys.\ Lett.\  B {\bf 279}, 377 (1992).
  %%CITATION = PHLTA,B279,377;%%
  %\bibitem{Luo:1994np}
  M.~Luo, J.~w.~Qiu and G.~Sterman,
  %``Anomalous nuclear enhancement in deeply inelastic scattering and
  %photoproduction,''
  Phys.\ Rev.\  D {\bf 50}, 1951 (1994).
  %%CITATION = PHRVA,D50,1951;%%
  %\bibitem{Luo:1993ui}
  M.~Luo, J.~W.~Qiu and G.~Sterman,
  %``Twist four nuclear parton distributions from photoproduction,''
  Phys.\ Rev.\  D {\bf 49}, 4493 (1994).
  %%CITATION = PHRVA,D49,4493;%%


 \bibitem{deltaE}
  %\bibitem{Zhang:2007ja}
  H.~Zhang, J.~F.~Owens, E.~Wang and X.~N.~Wang,
  %``Dihadron Tomography of High-Energy Nuclear Collisions in NLO pQCD,''
  Phys.\ Rev.\ Lett.\  {\bf 98}, 212301 (2007)
  %[arXiv:nucl-th/0701045].
  %%CITATION = PRLTA,98,212301;%%

 \bibitem{quadratically}
  %\bibitem{Baier:1996sk}
  R.~Baier, Y.~L.~Dokshitzer, A.~H.~Mueller, S.~Peigne and D.~Schiff,
  %``Radiative energy loss and p(T)-broadening of high energy partons in
  %nuclei,''
  Nucl.\ Phys.\  B {\bf 484}, 265 (1997)
  %[arXiv:hep-ph/9608322].
  %%CITATION = NUPHA,B484,265;%%

%  \bibitem{KKP}
%  %\bibitem{Kniehl:2000fe}
%  B.~A.~Kniehl, G.~Kramer and B.~Potter,
%  %``Fragmentation functions for pions, kaons, and protons at  next-to-leading
%  %order,''
%  Nucl.\ Phys.\  B {\bf 582}, 514 (2000)
%  [arXiv:hep-ph/0010289].
%  %%CITATION = NUPHA,B582,514;%%

 \bibitem{hadron-ratio}
 %\bibitem{Airapetian:2000ks}
  A.~Airapetian {\it et al.}  [HERMES Collaboration],
  %``Hadron formation in deep-inelastic positron scattering in a nuclear
  %environment,''
  Eur.\ Phys.\ J.\  C {\bf 20}, 479 (2001)
  %[arXiv:hep-ex/0012049].
  %%CITATION = EPHJA,C20,479;%%
  %\bibitem{Muccifora:2001zn}
  V.~Muccifora  [HERMES Collaboration],
  %``Hadron formation in DIS in a nuclear environment,''
  Nucl.\ Phys.\  A {\bf 715}, 506 (2003)
  %[arXiv:hep-ex/0106088].
  %%CITATION = NUPHA,A715,506;%%

 \bibitem{HERMES}
 %\bibitem{Airapetian:2007vu}
  A.~Airapetian {\it et al.}  [HERMES Collaboration],
  %``Hadronization in semi-inclusive deep-inelastic scattering on nuclei,''
  Nucl.\ Phys.\  B {\bf 780}, 1 (2007)
  %[arXiv:0704.3270 [hep-ex]].
  %%CITATION = NUPHA,B780,1;%%

 \bibitem{HKN}
  %\bibitem{Hirai:2007cx}
  M.~Hirai, S.~Kumano, T.~H.~Nagai and K.~Sudoh,
  %``Determination of fragmentation functions and their uncertainties,''
  Phys.\ Rev.\  D {\bf 75}, 094009 (2007)
  %[arXiv:hep-ph/0702250].
  %%CITATION = PHRVA,D75,094009;%%


 \bibitem{hadronic}
  B.~Z.~Kopeliovich,
  %``ARE HIGH-ENERGY QUARKS ABSORBED IN NUCLEAR MATTER?,''
  Phys.\ Lett.\  B {\bf 243}, 141 (1990).
  %%CITATION = PHLTA,B243,141;%%


\bibitem{Arleo:2003jz}
  F.~Arleo,
  %``Quenching of hadron spectra in DIS on nuclear targets,''
  Eur.\ Phys.\ J.\  C {\bf 30}, 213 (2003)
  %[arXiv:hep-ph/0306235].
  %%CITATION = EPHJA,C30,213;%%




\bibitem{Kharzeev:1996sq}
  D.~Kharzeev,
  %``Can Gluons Trace Baryon Number?,''
  Phys.\ Lett.\  B {\bf 378}, 238 (1996)
  %[arXiv:nucl-th/9602027].
  %%CITATION = PHLTA,B378,238;%%


\bibitem{Guo:1998rd}
  X.~F.~W.~Guo,
  %``Jet broadening in deeply inelastic scattering,''
  Phys.\ Rev.\  D {\bf 58}, 114033 (1998)
  %[arXiv:hep-ph/9804234].
  %%CITATION = PHRVA,D58,114033;%%


\bibitem{Guo:1999eh}
  X.~F.~Guo, J.~W.~Qiu and X.~F~Zhang,
  %``Nuclear dependence coefficient alpha(A,q(T)) for the Drell-Yan and  J/psi
  %production,''
  Phys.\ Rev.\  D {\bf 62}, 054008 (2000)
  %[arXiv:hep-ph/9912361].
  %%CITATION = PHRVA,D62,054008;%%


 \bibitem{dy}
 %\bibitem{McGaughey:1999mq}
  P.~L.~McGaughey, J.~M.~Moss and J.~C.~Peng,
  %``High-energy hadron-induced dilepton production from nucleons and  nuclei,''
  Ann.\ Rev.\ Nucl.\ Part.\ Sci.\  {\bf 49}, 217 (1999)
  %[arXiv:hep-ph/9905409].
  %%CITATION = ARNUA,49,217;%%



\end{thebibliography}
\end{document}